\documentclass[aps,prd,twocolumn,superscriptaddress,tightenlines,nofootinbib]{revtex4-1}

\usepackage{amsfonts,amsmath,amsthm,amssymb}
\usepackage{mathrsfs}
\usepackage{bm}
\usepackage{color}
\usepackage{epsfig}
\usepackage{mathrsfs}

\usepackage{delarray}
\usepackage{dcolumn}

\usepackage[usenames,dvipsnames]{xcolor}
\definecolor{orange}{cmyk}{0,0.5,1,0}
\definecolor{rossoCP3}{cmyk}{0,.88,.77,.40}
\definecolor{graa}{rgb}{0.8,0.8,0.8}
\definecolor{blaa}{rgb}{0.2,0.2,0.6}

\newcommand{\met} {\not\!\! E_T}

\newcommand{\beq}{\begin{equation}}
\newcommand{\eeq}{\end{equation}}
\newcommand{\bea}{\begin{flushleft} \begin{eqnarray}}
\newcommand{\eea}{\end{eqnarray}\end{flushleft}}

\newcommand{\postscript}[2]{\setlength{\epsfxsize}{#2\hsize}
   \centerline{\epsfbox{#1}}}
\newcommand{\comment}[1]{}

\newcommand{\er}[1]{\eqref{#1}}

\newcommand{\ci}[1]{}

\newcommand{\ke}{\rangle}
\newcommand{\br}{\langle}
\newcommand{\lb}{\left(}
\newcommand{\rb}{\right)}

\newcommand{\pd}{\partial}

\newcommand{\ba}{\begin{eqnarray}}
\newcommand{\ea}{\end{eqnarray}}
\newcommand{\be}{\begin{equation}}
\newcommand{\ee}{\end{equation}}
\newcommand{\bay}[1]{\left(\begin{array}{#1}}
\newcommand{\eay}{\end{array}\right)}

\newcommand{\ie}{\textit{i.e.}, }

\newcommand{\at}[1]{{|}_{#1}}

\def\met{\mbox{${\hbox{$E$\kern-0.6em\lower-.1ex\hbox{/}}}_T$}} 

\def\xl{{\lambda}}

\def\xt{{\theta}}

\def\CM{{\cal M}}

\newcommand{\beqa}{\begin{eqnarray}}
\newcommand{\eeqa}{\end{eqnarray}}

\newcommand{\la}{\langle}
\newcommand{\ra}{\rangle}
\newcommand{\lpa}{\left(}
\newcommand{\rpa}{\right)}

\begin{document}

\title{\color{rossoCP3}{ Majorana dark matter through the Higgs portal
    under the vacuum stability lamppost  
}}

\author{Luis A. Anchordoqui}
\affiliation{Department of Physics and Astronomy, Lehman College, City University of
  New York, NY 10468, USA
}

\affiliation{Department of Physics, Graduate Center, City University
  of New York, 365 Fifth Avenue, NY 10016, USA
}

\affiliation{Department of Astrophysics, American Museum of Natural History, Central Park West  79 St., NY 10024, USA}

\author {Vernon Barger}
\affiliation{Department of Physics, University of Wisconsin, Madison, WI 53706, USA}

\author{Haim \nolinebreak Goldberg}
\affiliation{Department of Physics,
Northeastern University, Boston, MA 02115, USA
}

\author{Xing \nolinebreak Huang}
\affiliation{Department of Physics, 
National Taiwan Normal University, Taipei, 116, Taiwan
}

\author{Danny Marfatia} 
\affiliation{Department of Physics and
  Astronomy, University of Hawaii, Honolulu, HI 96822, USA}

\author{Luiz H. M. da Silva}
\affiliation{Department of Physics,
University of Wisconsin-Milwaukee,
 Milwaukee, WI 53201, USA
}

\author{Thomas J. Weiler}
\affiliation{Department of Physics and Astronomy,
Vanderbilt University, Nashville TN 37235, USA
}

\date{June 2015}

\begin{abstract}
  \noindent We study the vacuum stability of a minimal Higgs portal
  model in which the standard model (SM) particle spectrum is extended
  to include one complex scalar field and one Dirac fermion. These new
  fields are singlets under the SM gauge group and are charged under a
  global $U(1)$ symmetry. Breaking of this $U(1)$ symmetry results in
  a massless Goldstone boson, a massive $CP$-even scalar, and splits
  the Dirac fermion into two new mass-eigenstates, corresponding to
  Majorana fermions. The lightest Majorana fermion $(w)$ is absolutely
  stable, providing a plausible dark matter (DM) candidate. We show
  that interactions between the Higgs sector and the lightest Majorana
  fermion which are strong enough to yield a thermal relic abundance
  consistent with observation can easily destabilize the electroweak
  vacuum or drive the theory into a non-perturbative regime at an
  energy scale well below the Planck mass. However, we also
  demonstrate that there is a region of the parameter space which
  develops a stable vacuum (up to the Planck scale), satisfies the
  relic abundance, and is in agreement with direct DM searches. Such
  an interesting region of the parameter space corresponds to DM
  masses $ 350~{\rm GeV} \alt m_w \alt 1 ~{\rm TeV}$. The region of
  interest is within reach of second generation DM direct detection
  experiments.

\end{abstract}

\maketitle

\section{Introduction}

The conspicuously well-known accomplishments of the $SU(3)_C \times
SU(2)_L \times U(1)_Y$ standard model (SM) of strong and electroweak
forces can be considered as the apotheosis of the gauge symmetry
principle to describe particle interactions. Most spectacularly, the
recent discovery~\cite{ATLAS:2012ae,Chatrchyan:2012tx} of a new boson
with scalar quantum numbers and couplings compatible with those of a
SM Higgs has possibly plugged the final remaining experimental hole in
the SM, cementing the theory further.

Arguably, the most challenging puzzle in high energy physics today is
to find out what is the underlying theory that completes the SM.  The
overly conservative approach to this dilemma has been to assess the
consistency of the SM assuming a vast desert between the electroweak
scale $M_{\rm EW} \sim 10^3~{\rm GeV}$ and the Planck mass $M_{\rm Pl}
\sim 10^{19}~{\rm GeV}$.  The relevant physics of the desert
hypothesis is determined by running couplings into the ultraviolet (UV)
using renormalization group (RG) equations. The behavior of the
running couplings depends sensitively on the weak scale boundary
conditions, among which the mass of the Higgs boson is perhaps the
most critical.  The measured Higgs mass $m_H = 125.5 \pm 0.5~{\rm
  GeV}$~\cite{Aad:2013wqa,Chatrchyan:2013mxa,Aad:2014aba,Khachatryan:2014ira}
corresponds to a Higgs quartic coupling $\lambda$ close to zero when
renormalized at energies above $\Lambda \sim 10^{11}~{\rm GeV}$.

Strictly speaking, next-to-leading order (NLO) constraints on SM
vacuum stability based on two-loop RG equations, one-loop threshold
corrections at the electroweak scale (possibly improved with two-loop
terms in the case of pure QCD corrections), and one-loop effective
potential seem to indicate $m_H$ saturates the minimum value that
ensures a vanishing Higgs quartic coupling around $M_{\rm Pl}$, see {\it
  e.g.}~\cite{Lindner:1988ww,Sher:1988mj,Diaz:1994bv,Casas:1994qy,Diaz:1995yv,Casas:1996aq,Isidori:2001bm,Isidori:2007vm,Hall:2009nd,Ellis:2009tp,EliasMiro:2011aa}. However,
the devil is in the details.  More recent NNLO
analyses~\cite{Bezrukov:2012sa,Degrassi:2012ry,Buttazzo:2013uya} yield a very
restrictive condition of absolute stability up to the Planck scale
\begin{eqnarray}
m_H & > & \left[129.4 + 1.4 \left( \frac{m_t/{\rm GeV} -173.1}{0.7}
\right)  \right. \nonumber \\
& - & \left. 0.5 \left(\frac{\alpha_s(m_Z) - 0.1184}{0.0007} \right) \pm
1.0_{\rm th} \right]~{\rm GeV}  \, .
\label{1}
\end{eqnarray}
On combining in quadrature the theoretical uncertainty with
experimental errors on the mass of the top ($m_t$) and the strong
coupling constant ($\alpha_s$), one obtains $m_H > 129 \pm 1.8~{\rm
  GeV}$. The vacuum stability of the SM up to the Planck scale is
excluded at 2$\sigma$ (98\% C.L. one sided) for $m_H < 126~{\rm
  GeV}$~\cite{Bezrukov:2012sa,Degrassi:2012ry,Buttazzo:2013uya}. 

The instability of the SM vacuum does not contradict any experimental
observation, provided its lifetime $\tau$ is longer than the age of
the universe $T_{\rm U}$.  Since the stability condition of the
electroweak vacuum is strongly sensitive to new physics, from the 
phenomenological point of view it is clear that beyond SM physics
models have to pass a sort of ``stability
test''~\cite{Branchina:2013jra,Branchina:2014rva,Lalak:2014qua}. Indeed, only new
physics models that reinforce the requirement of a stable or
metastable (but with $\tau > T_{\rm U}$) electroweak vacuum can be
accepted as a viable UV completion of the
SM~\cite{Basso:2010jm,Kadastik:2011aa,EliasMiro:2012ay,Cheung:2012nb,Anchordoqui:2012fq,Baek:2012uj,Chao:2012mx,Coriano:2014mpa,Altmannshofer:2014vra,Falkowski:2015iwa,Krog:2015cna,Rose:2015fua}.

From a theoretical perspective some modification of the Higgs sector
has long been expected, as the major motivation for physics beyond the
SM is aimed at resolving the huge disparity between the strength of
gravity and of the SM forces. Even if one abandons this hierarchy motivation, which does not
conflict with any experimental measurement, the SM has many other
(perhaps more basic) shortcomings. Roughly speaking, the SM is
incapable of explaining some well established observational
results. Among the most notable of these are neutrino masses, the QCD
theta parameter, and the presence of a large non-baryonic dark matter
(DM) component of the energy density in the universe. Interestingly,
if the new dynamics couples directly to the Higgs sector, this may
induce deviations from the usual vacuum stability and perturbativity
bounds of the SM.  However, beyond SM physics models are usually
driven by rather high scale dynamics ({\it e.g.}, the neutrino seesaw and
the QCD axion), in which case there will be a negligible effect on the
running of the couplings. A notable exception to this is the weakly
interacting massive particle (WIMP) DM, whose mass scale is
constrained to be low if produced by thermal freeze-out~\cite{Feng:2010gw}.

The scalar Higgs portal is a compelling model of WIMP DM in which a
renormalizable coupling with the Higgs boson provides the connection
between our visible world and a dark sector consisting of $SU(3)_C
\times SU(2)_L \times U(1)_Y$ singlet fields~\cite{Schabinger:2005ei,Patt:2006fw,Barger:2007im,Barger:2008jx}.
This is possible because the Higgs bilinear $\Phi^\dagger \Phi$ is the
only dimension-2 operator of the SM that is gauge and Lorentz
invariant, allowing for an interaction term with a complex singlet
scalar $S$ of the form 
\begin{equation}
\Delta V = \lambda_3 \Phi^\dagger \Phi
S^\dagger S \, . 
\end{equation} 
Given that $S$ develops a vacuum expectation value (VEV), the
Higgs mixes with the singlet leading to the existence of two mass
eigenstates ($h_1$ and $h_2$), which in turn open the portal into a
weak scale hidden sector. Despite its simplicity, in fact, this model
offers a rich phenomenology, and it provides a simple and motivated
paradigm of DM.

In this paper we carry out a general analysis of vacuum stability and
perturbativity in the SM augmented by a Higgs portal with a minimal
weak scale hidden sector. The layout is as follows. In Sec.~\ref{sec2}
we outline the basic setting of the scalar Higgs portal model and
discuss general aspects of the effective low energy theory resulting from a
minimal hidden sector.  In Sec.~\ref{sec3} we confront the model with
a variety of experimental data, including direct DM searches, heavy
meson decays with missing energy, the invisible Higgs width, as well
as astrophysical and cosmological observations. In Sec.~\ref{sec4}
we derive the RG equations and in Sec.~\ref{sec5} we present the analysis of vacuum
stability. Our conclusions are collected in Sec.~\ref{sec6}.

\section{Minimal Higgs Portal Model}
\label{sec2}

A viable DM candidate must be stable, or nearly so. Stability
results from either an unbroken or mildly broken symmetry in the
Lagrangian. A discrete $Z_2$ symmetry is the simplest available
symmetry to guarantee absolute stability of the DM particle.
Under $Z_2$ the SM particles are even while the DM particle
is odd~\cite{Krauss:1988zc}. The required symmetry may be simply
introduced by hand into the SM, or, more naturally, may remain after
breaking of some global continuous symmetry.  For example, a concrete
realization of such a hidden sector could emerge when a global $U(1)$
symmetry is spontaneously broken by a scalar field with charge 2 under
that symmetry, and so a discrete $Z_2$ symmetry arises automatically in
the Lagrangian. After spontaneous symmetry breaking, fields with an
even (odd) charge under the global $U(1)$ symmetry will acquire an
even (odd) discrete charge under $Z_2$.  Consequently the lightest
particle with odd charge will be absolutely stable, and thus a
plausible dark matter candidate.  The simplest approach to realize
this scenario is to introduce one new complex scalar field $S$ and one
Dirac fermion field $\psi$ into the SM.  These new fields are singlets
under the SM gauge group, and charged under $U(1)_W$ symmetry, such
that $U(1)_W (\psi) = 1$ and $U(1)_W(S) = 2$.  Spontaneous breaking of
a global continuous symmetry generates a massless Goldstone boson and
a $CP$-even scalar, and splits the Dirac fermion into two new
mass-eigenstates, corresponding to Majorana fermions.

The renormalizable scalar Lagrangian density of the set up described
above is found
to be
\begin{equation}
\label{new-scalar_L}
\mathscr{L}_s =\left( {\cal D}^{\mu} \Phi\right) ^{\dagger} {\cal
  D}_{\mu} \Phi + 
\left( {\cal D}^{\mu} S \right) ^{\dagger} {\cal D}_{\mu} S - V \, ,
\end{equation}
where 
\begin{equation}
V  =  \mu_1^2  \Phi^\dagger \Phi  +{ \mu_2}^2 S^\dagger 
S + \lambda_1  (\Phi^\dagger \Phi)^2  + \lambda_2 (S^\dagger S)^2 +
\Delta V 
\label{higgsV}
\end{equation}
is the potential and 
\beq {\cal D}_\mu = \partial_\mu - i g_2 \tau^a W^a_\mu - i g_Y Y
B_\mu 
\label{covderi2}
\eeq is (in a self-explanatory notation) the
covariant derivative. 
In the spirit of~\cite{Weinberg:2013kea}, we write $S$ in terms of two real fields (its massive
radial component and a massless Goldstone boson). The radial field
develops a VEV $\langle r \rangle$ about which the field $S$ is
expanded
\begin{equation}
S = \frac{1}{\sqrt{2}} \left(\langle r \rangle + r(x) \right) \ e^{i\, 2
    \alpha(x)} \, .
\end{equation}
The phase of $S$
is adjusted to make  $\langle \alpha (x) \rangle = 0$. 
Next, we impose the positivity conditions~\cite{Barger:2008jx}
\begin{equation}
\lambda_1 > 0, \quad \quad \lambda_2 > 0, \quad \quad
\lambda_1 \lambda_2 > \frac{1}{4} \lambda_3^2 \, .
\label{VernonGabePaul}
\end{equation}
If the conditions (\ref{VernonGabePaul}) are satisfied, we can proceed
to the minimization of (\ref{higgsV}) as a function of constant VEVs for
the two scalar fields.  In the unitary
gauge the Higgs doublet is expanded around the VEV as
\begin{equation}
\Phi(x) = \frac{1}{\sqrt{2}} \left( \begin{array}{c} 0 \\ \langle \phi
      \rangle + \phi(x) \end{array} \right) ,
\end{equation} 
where $\langle \phi \rangle = 246~{\rm GeV}$. 

The physically most
interesting solutions to the minimization of (\ref{higgsV}) are
obtained for $\langle \phi \rangle$ and $\langle r \rangle$ both non-vanishing
\begin{equation}
\langle \phi \rangle^2 = \frac{-\lambda _2 \mu_1^2 + \frac{1}{2} \lambda _3 {\mu_2}
  ^2}{\lambda _1 \lambda _2 - \frac{1}{4} \lambda _3^2}
\label{minima}
\end{equation}
 and
\begin{equation}
\langle r \rangle^2 = \frac{-\lambda _1 \mu_2^2 + \frac{1}{2} \lambda _3 \mu_1
  ^2}{\lambda _1 \lambda _2 - \frac{1}{4} \lambda _3^2} \, .
\label{ecuacionnueve}
\end{equation}
To compute the scalar masses, we must expand the potential
(\ref{higgsV}) around the minima (\ref{minima}) and
(\ref{ecuacionnueve}). We denote by $h_1$ and $h_2$ the scalar fields
of definite masses, $m_{h_1}$ and $m_{h_2}$ respectively.  After a bit
of algebra, the explicit expressions for the scalar mass eigenvalues
and eigenvectors read
\begin{equation}
m^2_{h_1} = \lambda _1 \langle \phi \rangle^2 + \lambda _2  \langle
r \rangle ^2 -
\zeta \, , \label{mh1}
\end{equation}
and
\begin{equation}
m^2_{h_2} = \lambda _1 \langle \phi \rangle^2 + \lambda _2  \langle
r \rangle^2  + \zeta \, ,
 \label{mh2}
\end{equation}
with
\begin{equation}
\zeta = \left| \sqrt{(\lambda _1
  \langle \phi \rangle^2 - \lambda _2   \langle r \rangle^2)^2 + (\lambda _3 \langle
  \phi \rangle \langle r \rangle)^2} \right|
\end{equation}
and
\begin{equation}
\left(\begin{array}{c} h_1 \\ h_2 \end{array} \right) = \left( \begin{array}{cc} 
\cos \theta & - \sin \theta \\ \sin \theta & \phantom{-} \cos \theta \end{array} \right) 
\left(\begin{array}{c} r \\ \phi \end{array} \right) \, . 
\end{equation}
Here, $\theta \in [-\pi/2, \pi/2]$ also  fullfils
\begin{equation}\label{sin2a}
\sin{2\theta} = \frac{\lambda _3 \langle \phi \rangle
  \langle r \rangle}{\sqrt{(\lambda _1 \langle \phi \rangle^2 - \lambda _2 \langle r \rangle^2)^2
    + (\lambda _3 \langle \phi \rangle \langle r \rangle)^2}} \, .
\end{equation}
Now, it is convenient to invert (\ref{mh1}), (\ref{mh2}) and (\ref{sin2a}), 
to extract the parameters in the Lagrangian in terms of  measurable
quantities: $m_{h_1}$, $m_{h_2}$ and $\sin{2\theta}$. We obtain 
\begin{eqnarray}
\label{12}
\lambda _1 &=& \frac{m_{h_{2,1}}^2}{4\langle \phi \rangle^2}(1-\cos{2\theta}) +
\frac{m_{h_{1,2}}^2}{4\langle \phi \rangle^2}(1+\cos{2\theta}), \nonumber \\ 
\lambda _2 &=& \frac{m_{h_{1,2}}^2}{4\langle r \rangle^2}(1-\cos{2\theta}) +
\frac{m_{h_{2,1}}^2}{4 \langle r \rangle^2}(1+\cos{2\theta}),\\ 
\lambda _3 &=& \sin{2\theta} \left( \frac{m_{h_{2,1}}^2-m_{h_{1,2}}^2}{2\langle
    \phi \rangle \langle r \rangle}
\right). \nonumber
\end{eqnarray}
Note that there are two distinct regions of the parameter space: one
in which the hidden scalar singlet is heavier than the Higgs doublet
and one in which is lighter. The small $\theta$ limit leads to the
usual SM phenomenology with an isolated hidden sector.

For  the DM sector we assume at least one Dirac field
\beq
\mathscr{L}_\psi = i \bar{\psi}\gamma \cdot \pd \psi - m_\psi \bar{\psi} \psi 
- \frac{f}{\sqrt{2}} \bar{\psi^c} \psi \, S^\dagger  -
\frac{f^*}{\sqrt{2}} \bar{\psi} \psi^c \, S \,  .
\label{eq:fortyone}
\eeq 
As advanced above,
we assign to the hidden fermion a charge $U(1)_W (\psi) = 1$, so that the Lagrangian is
invariant under the global transformation $e^{i W \alpha}$. Assuming the
 transformation is local we express $\psi$ as \beq
\psi(x) = \psi'(x) e^{i \alpha(x)}.
\label{eq:fortytwo}
\eeq
Now, after $r$ achieves a VEV we expand the DM sector to obtain \beqa \mathscr{L}_\psi &=&
\frac{i}{2}\left(\bar{\psi}'\gamma \cdot \pd \psi' + \bar{\psi'}^{c}
  \gamma \cdot \pd \psi^{c'} \right), \nonumber
\\
&-& \frac{m_\psi}{2} \left( \bar{\psi}' \psi' + \bar{\psi'}^{c}
  {\psi'}^{c} \right)-\frac{f \langle r \rangle}{2} \bar{\psi'}^{c} \psi' - \frac{f
  \langle r \rangle}{2} \bar{\psi}' {\psi'}^{c} , \nonumber
\\
&-& \frac{1}{2} (\bar{\psi}' \gamma \psi' - \bar{\psi'}^{c} \gamma
{\psi '}^{c} ) \cdot \pd \alpha , \nonumber
\\
&-& \frac{f}{2} r \left( \bar{\psi'}^{c}\psi' + \bar{\psi}' {\psi'}^{c}
\right).
\label{eq:fortyfour}
\eeqa
The diagonalization of the $\psi'$ mass matrix  generates the mass eigenvalues,
\begin{equation}
m_\pm =  m_\psi \pm  f  \langle r \rangle, 
\end{equation}
for the two mass eigenstates
\begin{equation}
\psi_- = \frac{i}{\sqrt{2}} \lpa \psi'^c - \psi'  \rpa
  \quad {\rm and}   \quad \psi_+ = \frac{1}{\sqrt{2}}\lpa \psi'^c+\psi' \rpa  \, .
\label{eq:fourtyseven}
\end{equation}
In the new basis, the act of charge conjugation on $\psi_\pm$ yields
\beq
\psi^c_\pm =  \psi_\pm \, ,
\label{eq:fourtyeight}
\eeq 
which implies that the fields $\psi_\pm$ are Majorana
fermions. The Lagrangian is found to be 
\beqa \mathscr{L}_\psi
&=&\frac{i}{2}\bar{\psi_+}\gamma \cdot \pd \psi_+ +
\frac{i}{2}\bar{\psi_-}\gamma \cdot \pd \psi_- \nonumber \\
& - & \frac{1}{2} m_+
\bar{\psi}_+ \psi_+ - \frac{1}{2}m_- \bar{\psi}_- \psi_- , \nonumber
\\
&-&\frac{i}{4 \langle r \rangle} (\bar{\psi}_+ \gamma \psi_- + \bar{\psi}_-
\gamma \psi_+) \cdot \pd \alpha' , \nonumber
\\
& -& \frac{f}{2} r (\bar{\psi}_+\psi_+ + \bar{\psi}_- \psi_-) \,,
\label{eq:fortynine} 
\eeqa where $\alpha' \equiv 2 \alpha \langle r \rangle$ is the canonically normalized
Goldstone boson~\cite{Weinberg:2013kea}.  We must now put $r$ into its massive field
representation, for which the interactions of interest are 
\begin{eqnarray}
\mathscr{L} & = & -\frac{f \sin \theta}{2} h_{1,2} (\bar{\psi}_+\psi_+ + \bar{\psi}_- \psi_-) -
\frac{f \cos \theta}{2} h_{2,1} \nonumber \\
& \times & (\bar{\psi}_+\psi_+ + \bar{\psi}_- \psi_-).
\label{eq:fifty}
\end{eqnarray}
This leads to 3-point interactions between the Majorana fermions and the
Higgs doublet.

\begin{table}
\caption{Definition of most common variables. \label{table:0}}
\begin{tabular}{ll}
\hline
\hline
$\Phi$  & ~~Higgs doublet \\
$S$  & ~~Complex scalar field \\
$\phi$ & ~~Neutral component of $\Phi$ \\
$r$ & ~~Massive $CP$-even scalar \\
$\alpha'$ & ~~Goldstone boson \\
$H$ & ~~SM Higgs boson \\
$h_{1,2}$ & ~~Scalar mass eigenstates \\
$\lambda_3$ & ~~Quartic  coupling between SM and  hidden sector  \\
$\theta$ & ~~Mixing angle between $h_1$ and $h_2$ \\
$w$ & ~~Lightest Majorana fermion (WIMP) \\
$f$ & ~~$w-r$ coupling constant -- see Eq.~(\ref{eq:fortynine}) -- \\
\hline
\hline
\end{tabular}
\end{table}

All in all, the Dirac fermion of the hidden sector splits into two
Majorana mass-eigenstates. The heavier state will decay into the
lighter one by emitting a Goldstone boson. The lighter one, however,
is kept stable by the unbroken reflection symmetry. Hence, we can
predict that today the universe will contain only one species of
Majorana WIMP, the lighter one $w$, with mass $m_w$ equal to the
smaller of $m_\pm$. Therefore, the dark sector contains five unknown
parameters: $m_w$, $m_{h_{1,2}}$, $\lambda_2$, $\theta$, and $f$. To
facilitate the calculation of the WIMP relic density, throughout we
impose a supplementary constraint relating some of these free
parameters: $\Delta m/m_w \ll 1$, where \mbox{$\Delta m = |m_+ - m_-|
  = 2 |f \langle r \rangle|$.}  (The most common variables used in
this article are summarized in Table~\ref{table:0}.)

A cautionary note is worth taking on board at  this juncture. It is
well known that the
spontaneous breaking of a global $U(1)$ symmetry have several
disconnected and degenerate vacua (the phase of the vacuum expectation
value $\langle 0 | S |0 \rangle$ can be different in different
regions of space, and actually we expect it to be different in
casually disconnected regions), yielding dangerous domain-wall
structure in the early
universe~\cite{Sikivie:1982qv,Vilenkin:1982ks}. In the spirit
of~\cite{Sikivie:1982qv}, it may be possible to explicitly break the symmetry 
introducing (possibly small) terms in $V$, such that the domain walls
disappear before dominating the matter density of the universe, while
leaving \mbox{(pseudo-)Goldstone} bosons and the same dark matter
phenomenology~\cite{EliasMiro:2012ay}.\footnote{Other approaches, if exceedingly fine-tuned,
  may offer alternative solutions~\cite{Turner:1990uz,Dvali:1991ka,Hiramatsu:2012sc}.}
For simplicity, we restrict our considerations to the potential in  (\ref{higgsV}), but generalizations are straightforward.

\section{Constraints from experiment}
\label{sec3}

The mixing of $r$ with the Higgs doublet $\phi$ can be analyzed in a
two-parameter space characterized by the mass of hidden scalar
$m_{h_i}$ and the mixing angle $\theta$, where $i = 1$ for a light
scalar singlet ({\it i.e.} $m_{h_2} = m_H$) and $i=2$ for a heavy one
({\it i.e.} $m_{h_1} = m_H$). We begin to constrain this parameter
space by using data from DM searches at direct detection
experiments. 

\subsection{Constraints from direct DM searches}

The $wN$ cross section for elastic scattering is  
found to be
\begin{equation}
\sigma_{wN} = \frac{4}{\pi} \frac{m_w^2 m_N^2}{(m_w + m_N)^2} \ \frac{f_p^2 + f_n^2 }{2} \,,
\label{sigmawN}
\end{equation}
where $N \equiv \frac{1}{2} (n+p)$ is an isoscalar nucleon in the
renormalization group-improved parton
model~\cite{Ellis:2000ds,Beltran:2008xg}. The effective couplings to
protons $f_p$ and neutrons $f_n$ are given by 
\begin{eqnarray} f_{p,n} & = & \sum_{q =
  u,d,s} \frac{G_q}{\sqrt{2}} f_{Sq}^{(p,n)} \frac{m_{p,n}}{m_q} +
\frac{2}{27} f_{SG}^{(p,n)} \nonumber \\
& \times & \sum_{q = c,b,t} \frac{G_q}{\sqrt{2}}
\frac{m_{p,n}}{m_q},
\label{effective-coup}
\end{eqnarray}
where $G_q$ is the WIMP's effective Fermi coupling for a
given quark species, \beq \mathscr{L} = \frac{G_{q}}{\sqrt{2}} \bar
\psi_- \psi_- \bar \psi_q \psi_q \,, \eeq with $\psi_q$ the SM quark
field of flavor $q$. The first term in (\ref{effective-coup}) reflects
scattering with light quarks, whereas the second term accounts for
interaction with gluons through a heavy quark loop. The scalar
spin-independent form factors,
$f_{Sq}^{(p,n)}$, are proportional to the matrix element, $\la \bar q
q \ra$, of quarks in a nucleon. Herein we take~\cite{Hill:2014yxa}
\beqa
f^{p}_{Su} = 0.016(5) (3) (1),  & \quad \ f^{n}_{Su} =
0.014(5)(^{+2}_{-3}) (1),
\nonumber \\ f^{p}_{Sd} = 0.029 (9) (3) (2) , &\quad
f^{n}_{Sd} = 0.034(9) (^{+3}_{-2}) (2), \nonumber \\  f^{p}_{Ss} = 0.043(21),~~~~~~ &
\quad f^{n}_{Ss} = 0.043 (21) \, , ~~~~~~~~~ \eeqa 
in good agreement with the scalar strange content of the nucleon from
lattice QCD calculations~\cite{Junnarkar:2013ac}.
The gluon scalar form factor is given by $f^{(p,n)}_{SG} = 1 - \sum_{u,d,s} f^{(p,n)}_{Sq}$.
For the case at hand,
\begin{equation}
\frac{f_p^2 +f_n^2}{2 m_N^2} \simeq  \left(0.29 \frac{G_q}{\sqrt{2} m_q} \right)^2, 
\end{equation}
with
\begin{equation}
 \frac{G_q}{m_q} = \frac{\sqrt{2} f \lambda_3
\langle r \rangle}{2  m_{h_1}^2 m_{h_2}^2}, 
\end{equation}
yielding~\cite{Anchordoqui:2013pta} 
\beq
\sigma_{wN} =  \frac{1}{\pi}
\frac{m_w^2 m_N^4}{(m_w + m_N)^2} \left( \frac{0.29\, \lambda_3 \, \langle r
    \rangle \, f}{ m_{h_1}^2 m_{h_2}^2} \right)^2 \ ;
\eeq
see Appendix~\ref{AA} for details.
We may re-express this result in terms of the mixing angle,
\begin{eqnarray}
\sigma_{wN} & = & (0.29)^{2} \ \frac{1}{4\pi} \  \frac{m_w^2 m_N^4}{(m_w +
  m_N)^2} \ \left(
  \frac{1}{m_{h_1}^2} - \frac{1}{m_{h_2}^2} \right)^2 \nonumber \\
& \times & \left( \frac{ f}{ \langle \phi \rangle} \right)^2  \ \sin^2 2 \theta \ .
\end{eqnarray}
For $\theta \ll 1$, the upper limits on the nucleon-wimp cross
sections derived by the various experiments translate into upper
limits on the mixing angle
\begin{eqnarray}
|\theta| & < & \frac{(m_w + m_N)}{m_N^2 m_w} \frac{\langle \phi
  \rangle}{f}  \left| \frac{1}{m_{h_1}^2} - \frac{1}{m_{h_2}^2} \right|^{-1}
\nonumber \\
& \times &
\frac{\sqrt{\pi}}{0.29} \sqrt{ \sigma_{wN}(m_w) }  \ .
\label{DM_bound}
\end{eqnarray}

\begin{figure}[tbp]
\postscript{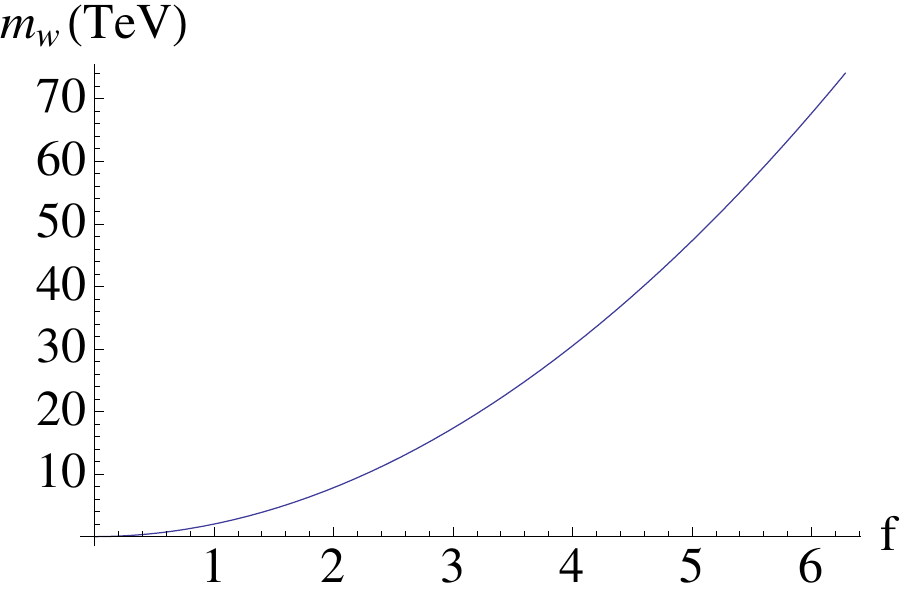}{1}
\caption{The relation in Eq.~(\ref{efe}).}
\label{rg-w_f1}
\end{figure}

To determine $f$ we require the $w$ relic density to be consistent
with $h^2 \Omega_{\rm DM} \simeq 0.111(6)$~\cite{Agashe:2014kda}.  In our study we consider
the interesting case in which $m_{h_i} < m_w$ and hence the
instantaneous freeze-out approximation is valid~\cite{Garcia-Cely:2013nin}.  In this region of
the parameter space, the $w$'s predominantly annihilate into a pair of $h_i$'s or co-annihilate with the next-to-lightest
Majorana fermion, producing a scalar $h_i$ and a Goldstone boson.  All
of the final state $h_i$ subsequently decays into $\alpha'$. We note,
however, that for $m_w \approx m_H/2$ one expects dominant
annihilation into fermions.  We have found that for the considerations
in the present work, the effective thermal cross section can be safely
approximated by~\cite{Garcia-Cely:2013nin}  
\begin{equation}
\lim_{\Delta m /m_w
  \rightarrow 0} \la \sigma_{ww} v_M \ra \approx \frac{f^4}{32 \pi
  m_w^2}, 
\end{equation}
yielding
\begin{equation}
f \approx \left( \frac{1.04 \times 10^{11}~{\rm GeV^{-1}} \, x_f}{\sqrt{g(x_f)} \ M_{\rm Pl} \, \Omega_{\rm DM} h^2} \right)^{1/4} \sqrt{m_w}  \ ,
\label{efe}
\end{equation}
where $x_f = m_w/T_f$, $g(x_f)$ is is the number of relativistic
degrees of freedom at the freeze-out temperature $T_f$, and
$m_{h_i}/m_w \alt 0.8$~\cite{Garcia-Cely:2013nin}. In general for WIMP
DM $x_f \approx 20 - 25$~\cite{Gondolo:1990dk}.  The precise relation
between the WIMP mass and the required Yukawa coupling to attain the
relic density condition is shown in Fig.~\ref{rg-w_f1}.  We note that
the mass upper limit, $m_w < 74~{\rm TeV}$, is in agreement with the
unitarity limit $\Omega_{\rm DM} h^2 \geq 1.7 \times 10^{-6}
\sqrt{x_f} \left[m_w/(1~{\rm TeV})\right]^2$~\cite{Griest:1989wd},
which implies $m_w \leq 110~{\rm TeV}$~\cite{Blum:2014dca}.

\begin{figure*}[tbp]
\begin{minipage}[t]{0.49\textwidth}
\postscript{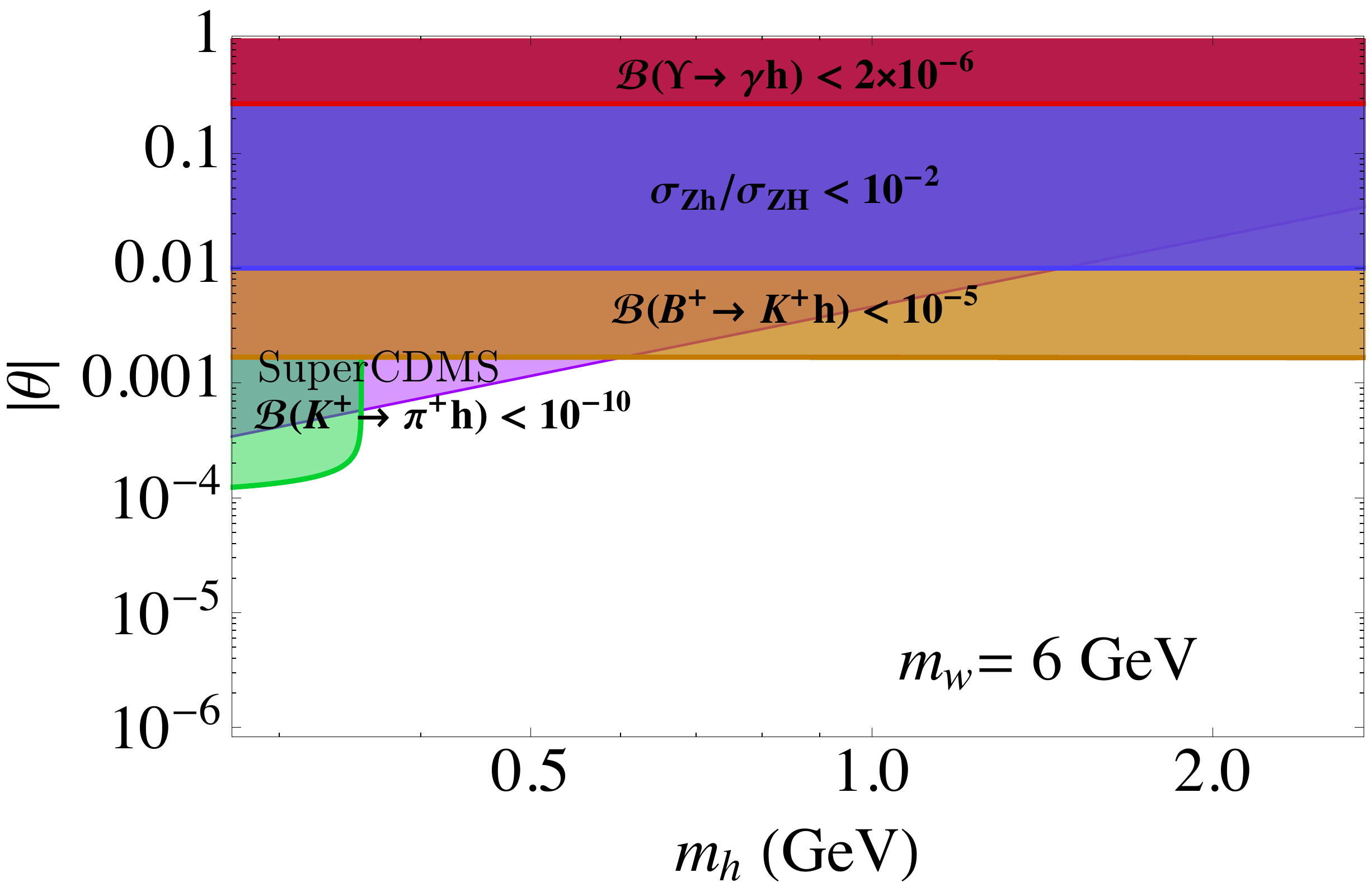}{0.99}
\end{minipage}
\hfill
\begin{minipage}[t]{0.49\textwidth}
\postscript{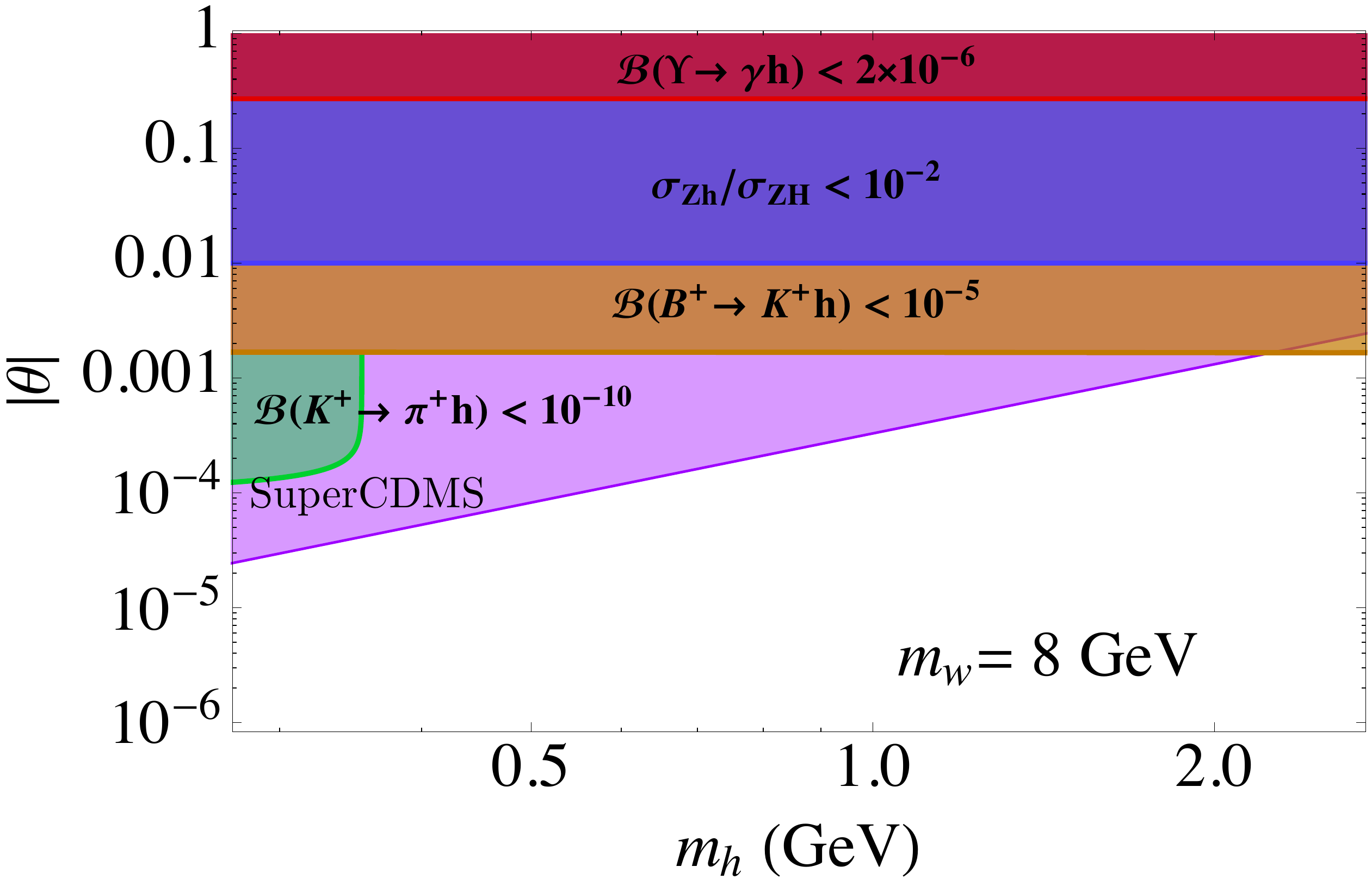}{0.99}
\end{minipage}
\begin{minipage}[t]{0.49\textwidth}
\postscript{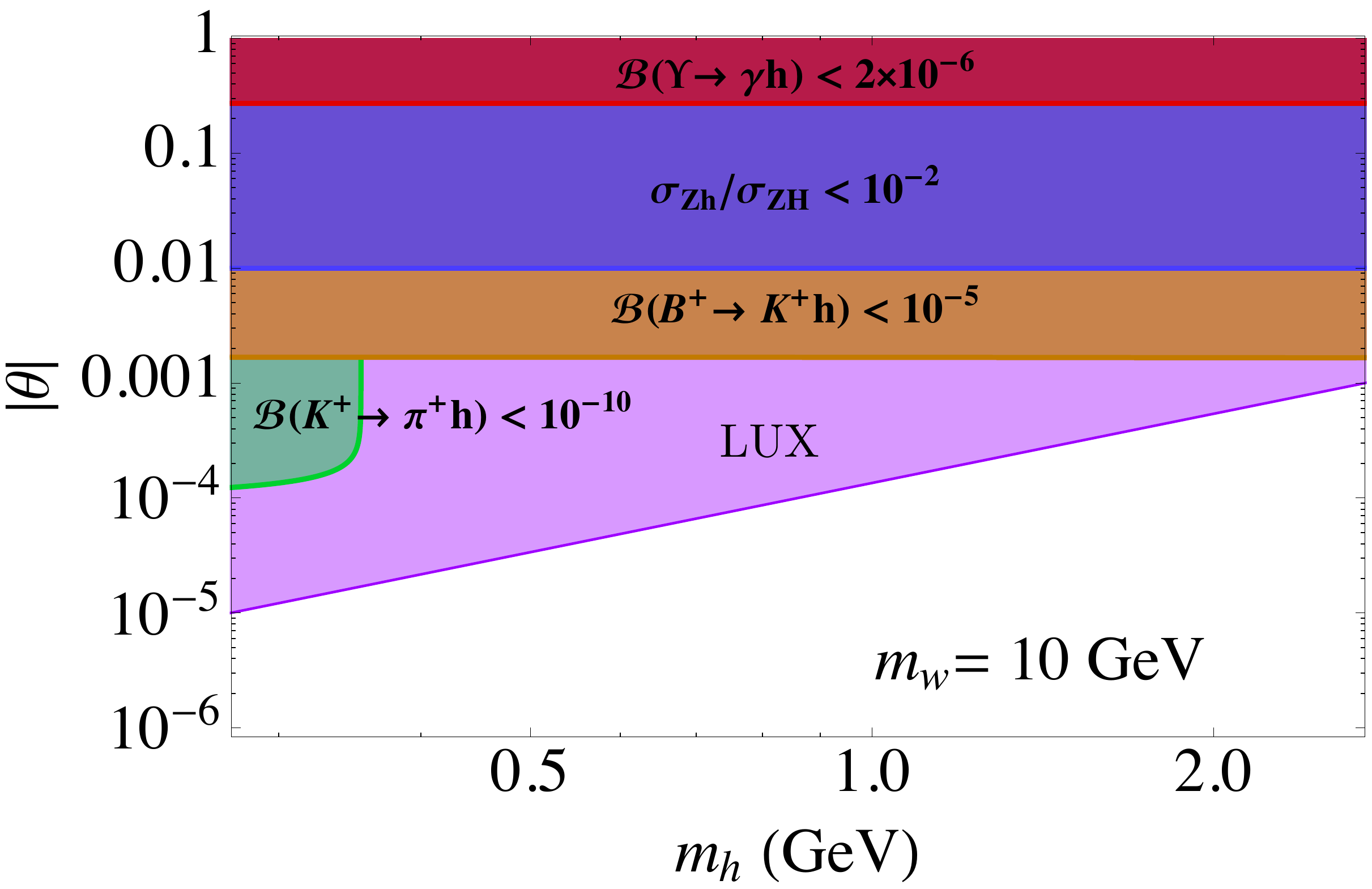}{0.99}
\end{minipage}
\hfill
\begin{minipage}[t]{0.49\textwidth}
\postscript{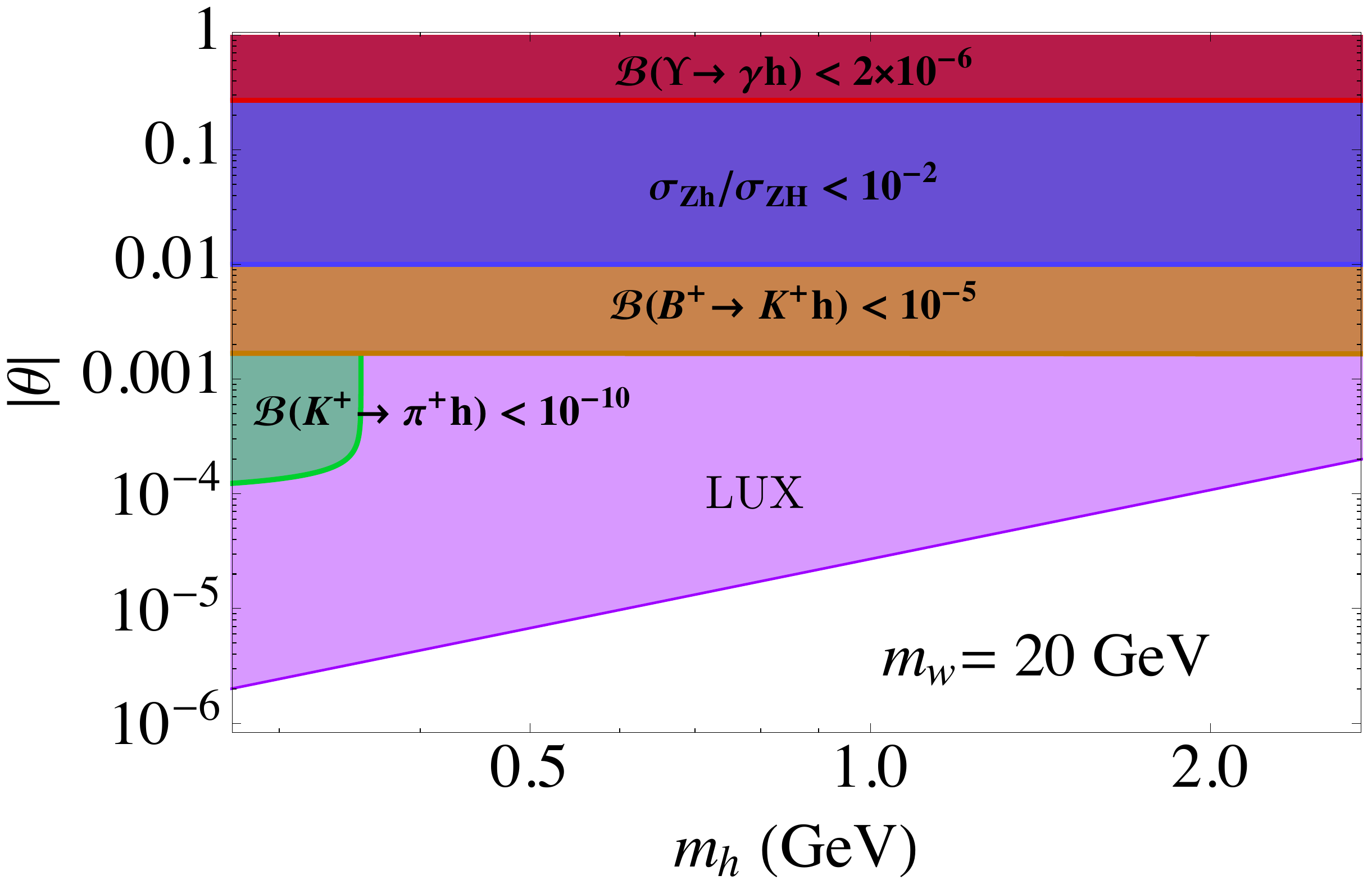}{0.99}
\end{minipage}
\begin{minipage}[t]{0.49\textwidth}
\postscript{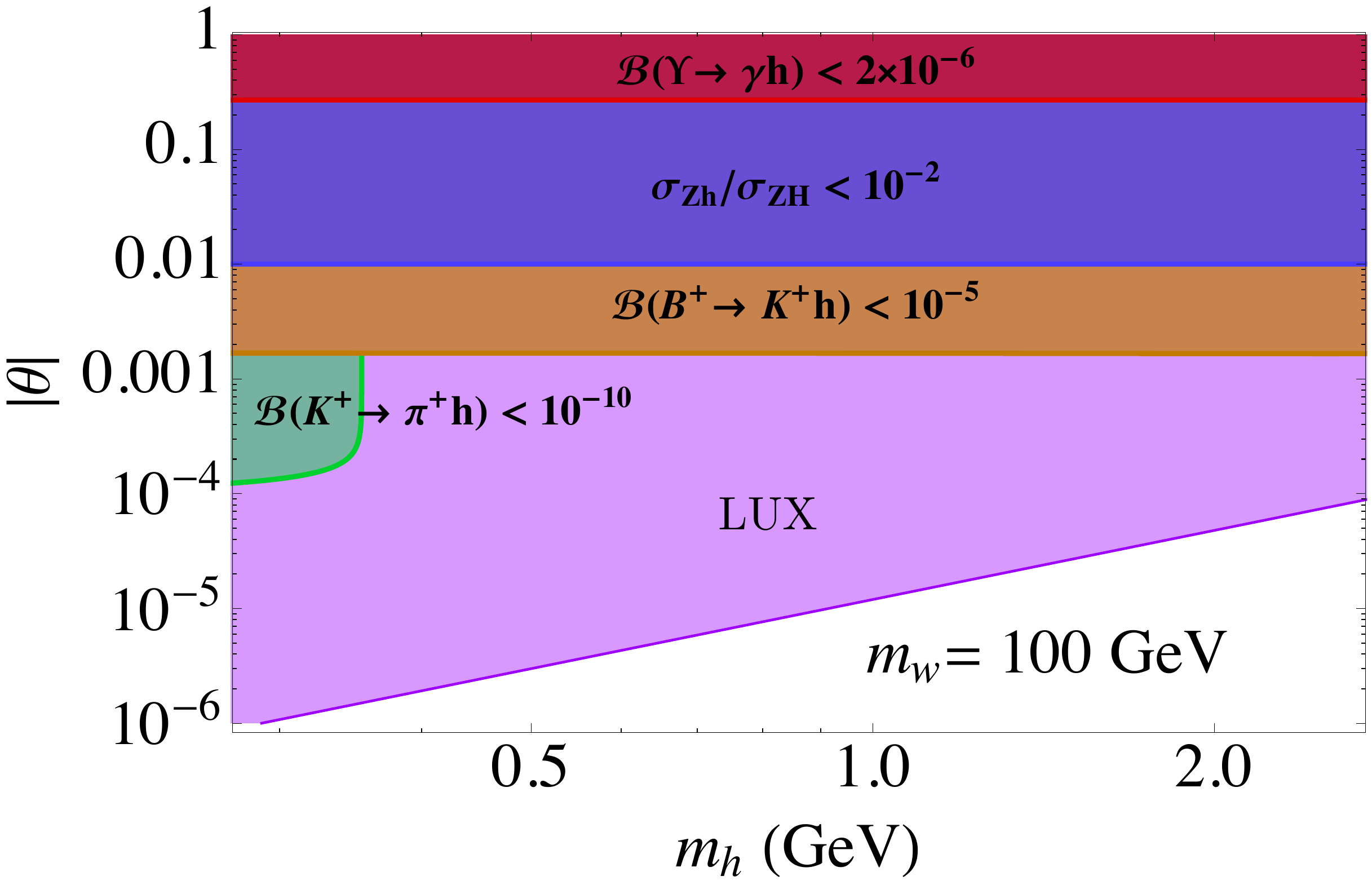}{0.99}
\end{minipage}
\hfill
\begin{minipage}[t]{0.49\textwidth}
\postscript{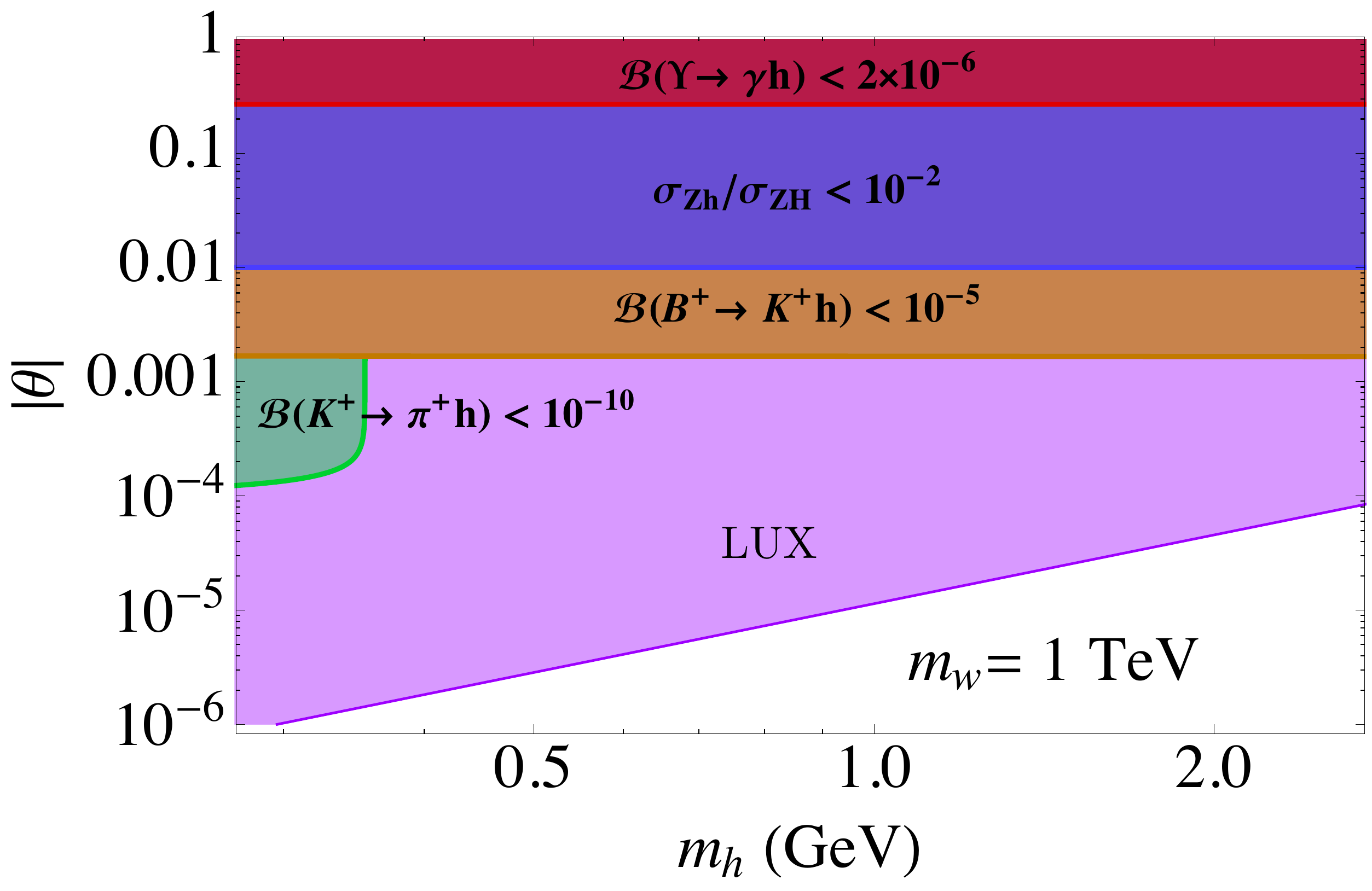}{0.99}
\end{minipage}
\caption{Excluded regions of the ($|\theta|, m_h$) parameter space from interactions involving SM particles in
the initial state and the $CP$-even scalar in the final state, as well from DM direct detection
experiments. The horizontal bands indicate bounds are from heavy meson
decays with missing energy  (no significant excess of such decays over
background has been observed yielding bounds on the processes $\Upsilon \to \gamma h, \  B^+ \to K^+ h, \ K^+
\to \pi^+ h$) as well as from
LEP limits on the production of invisibly-decaying 
Higgs bosons $\sigma_{Zh}/\sigma_{ZH}$. The diagonal bands represent bounds from DM direct detection
experiments (Super-CDMS and LUX), for different values of the WIMP
mass. Note that all bounds other than the LEP bound can be smoothly
extrapolated to the smallest $m_h \sim 35~{\rm MeV}$ stipulated by cosmology.}
\label{rg-w_f2}
\end{figure*}

Using (\ref{DM_bound}) we can now translate the 90\% confidence limit
on the spin independent elastic WIMP-nucleon cross section as obtained
by direct detection experiments into an upper limit of $|\theta|$. In
Fig.~\ref{rg-w_f2} we show constraints on this parameter space from
direct dark matter searches. For $m_w \agt 8~{\rm GeV}$, the most
restrictive constraint comes from the LUX
experiment~\cite{Akerib:2013tjd}, whereas for $m_w \alt 8~{\rm GeV}$,
the most restrictive upper limit is from the SuperCDMS low threshold
experiment~\cite{Agnese:2014aze}. It should be noted that
indirect DM searches ({\it e.g.} by detecting neutrinos from
annihilation of captured low-mass WIMPs in the Sun) also
constrain the WIMP-nucleon elastic scattering cross section. However, these searches are in general
model dependent. For example, for 100\% annihilation into $\tau^+
\tau^-$ pairs, the Super-Kamiokande Collaboration~\cite{Choi:2015ara}
has set the current best upper limit on  $\sigma_{wN}$ for WIMP masses below $8~{\rm
  GeV}$. Because of the assumed dominant decay into SM fields, this
limit cannot be used to further constrain the $(\theta, m_h)$ parameter space. 

\subsection{Constraints from heavy meson decay}

For $m_w \alt 10~{\rm GeV}$, searches for heavy meson decays with
missing energy provide comparable
bounds~\cite{Schmidt-Hoberg:2013hba,Clarke:2013aya,Anchordoqui:2013bfa}. In
particular, the upper limit reported by the BaBar Collaboration
\mbox{${\cal B} (\Upsilon (1S) \to \gamma + \met) < 2 \times
  10^{-6}$~\cite{delAmoSanchez:2010ac}} yields an upper bound for the
mixing angle, $\theta <
0.27$~\cite{Huang:2013oua}.\footnote{Comparable bounds are obtained
  from searches for ${\cal B} (\Upsilon(3S) \to \gamma +
  \met)$~\cite{Aubert:2008as} and ${\cal B} (J/\psi \to \gamma +
  \met)$~\cite{Insler:2010jw}.}  A stronger constraint follows from
LEP limits on the production of invisibly-decaying Higgs bosons
$\sigma_{Zh}/\sigma_{ZH} <
10^{-4}$~\cite{Buskulic:1993gi,Acciarri:1996um,Alexander:1996ne,Abbiendi:2002qp,Abbiendi:2007ac},
which implies $\theta < 10^{-2}$~\cite{Cheung:2013oya}. More
restrictive constraints come from searches for the rare
flavor-changing neutral-current decay $B^+ \rightarrow K^+ + \met$
reported by the
BaBar~\cite{delAmoSanchez:2010bk,Lees:2013kla,Aubert:2004ws},
CLEO~\cite{Browder:2000qr}, and BELLE~\cite{Lutz:2013ftz}
collaborations, as well as limits on $K^+ \to \pi^+ + \met$ from the
E787~\cite{Adler:2001xv} and E949
experiments~\cite{Anisimovsky:2004hr,Adler:2008zza,Artamonov:2009sz}. The
resulting excluded regions of the ($|\theta|, m_h$) plane from all
these experiments are compared in Fig.~\ref{rg-w_f2} with those from
direct DM searches.

\subsection{Constraints from LHC and SN1987A}

\begin{figure}[tbp]
\postscript{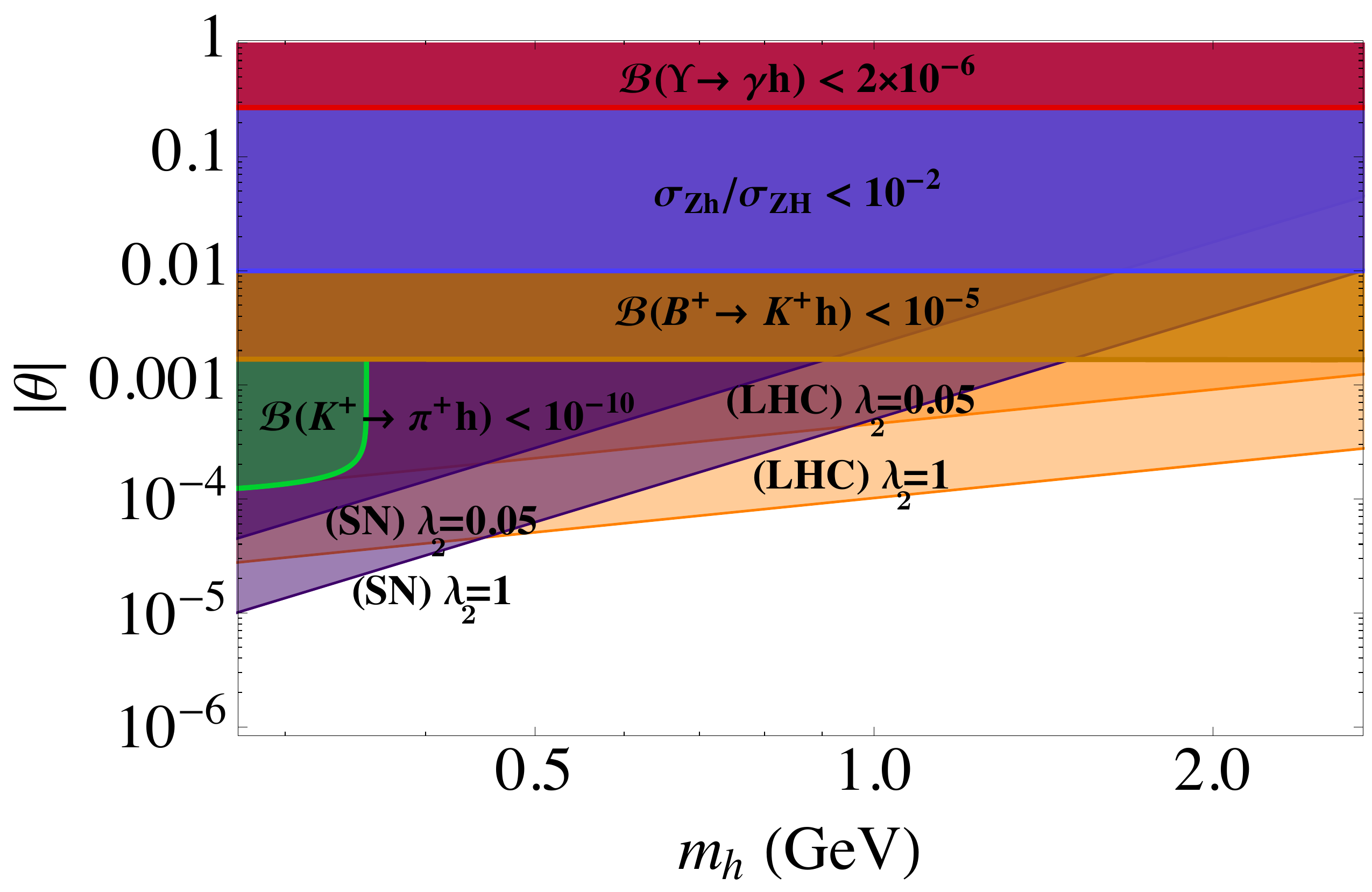}{1}
\caption{Bounds on the $(|\theta|, m_h)$ including invisible Higgs
  decays and $\alpha'$ emission in a post-collapse supernova core  for different assumptions about the value of the quartic 
coupling $\lambda_2$.}
\label{rg-w_f3}
\end{figure}

Before proceeding we note that additional constraints on the $(|\theta|,
m_h)$ parameter space can be obtained from limits on Higgs decay into
invisible particles and from emission of $\alpha'$-particle pairs in a post-collapse
supernova core.  However, these are not direct constraints as they depend also on the quartic coupling of the hidden scalar. In particular,
since invisible decays reduce the branching fraction to the (visible)
SM final states, it is to be expected that ${\cal B} (H \to \, {\rm
  invisible})$ is strongly constrained. Indeed ${\cal B} (H \to \,
{\rm invisible})$ is known to be less than about 19\% at
95\%CL~\cite{Barger:2012hv,Espinosa:2012vu,Cheung:2013kla,Giardino:2013bma,Ellis:2013lra}.
This implies exclusion
contours in the $(|\theta|, m_h)$ plane as a function of the free
parameter $\lambda_2$ given by~\cite{Anchordoqui:2013bfa} 
$$
|\theta (\lambda_2)|  <  1.27 \times 10^{-2}  \left[ \lambda_2
  \frac{m_H^2}{m_h^2} +  f^2 \sqrt{1 - \frac{4 m_w^2}{m_H^2}} \,
\right]^{-\frac{1}{2}}. 
$$
In addition, the emissivity of $\alpha'$ due to nucleon
bremsstrahlung ($NN \rightarrow NN \alpha' \alpha'$) cannot exceed the
limits imposed by SN1987A observations: $\epsilon_{\alpha'} \leq 7.324
\times 10^{-27} \ {\rm GeV}$~\cite{Raffelt:1990yz}. For typical
supernova core conditions ($T = 30~{\rm MeV}$ and $\rho = 3 \times
10^{14}~{\rm g/cm^3}$) it is easily seen that $|\lambda_3| \leq 0.011
\left( \frac{m_h}{500 \ {\rm MeV}} \right)^2$~\cite{Keung:2013mfa}.
For $\theta \ll 1$ we can translate this limit into a bound on the
mixing angle via \beq \theta \approx \frac{\lambda_3 \ \la r \ra \la
  \phi \ra}{m_H^2 - m_h^2} \ .  \eeq By use of $m_h \approx \sqrt{2
  \lambda_2} \la r \ra$ we can express this bound as \beq |\theta|
\leq \frac{7.65 \ m_h^3}{\sqrt{\lambda_2} |m_H^2 - m_h^2| } \ {\rm
  GeV}^{-1} \ .  
\label{sn1987}
\eeq 
In Fig.~\ref{rg-w_f3} we show the exclusion contours for the
$\lambda_2 = 1$ and $\lambda_2 = 0.05$. For smaller values of
$\lambda_2$, the excluded regions of the $(|\theta|,m_h)$ plane are
dominated by upper limits on
$B$-meson decay into invisibles. All in all, for $m_{h_2} = m_H$, we can conclude from
Figs.~\ref{rg-w_f2} and \ref{rg-w_f3} that $2 \times 10^{-3}$ is a conservative 90\% CL
upper limit on the mixing angle.

For $m_{h_2} \gg m_H$, (\ref{DM_bound})  can be rewritten as
\begin{eqnarray}
f \ |\theta|  & < & \frac{1}{m_N^2} \ \langle \phi
  \rangle  \  m_H^2 \ 
\frac{\sqrt{\pi}}{0.29} \ \sqrt{ \sigma_{wN}(m_w) }  \nonumber \\
& \simeq & 2.7 \times 10^{7}  \ \sqrt{\sigma_wN (m_w)}~{\rm GeV} \, .
\label{DM_bound2}
\end{eqnarray}

Dedicated searches for DM candidates serve as an essential component
of the LHC physics programme. The typical experimental signature of DM
production at the LHC consists of an excess of events with a single
final-state partilce $X$ recoiling against large amounts of missing
transverse momentum or energy. In Run I, the ATLAS and CMS
collaborations have examined a variety of such ``mono-$X$'' topologies
involving jets of hadrons, gauge bosons, top and bottom quarks as well
as the Higgs boson in the final state. In particular, the CMS
Collaboration has reported very restrictive bounds on the DM-nucleon
scattering cross section from searches in events containing a jet and
an imbalanced transverse momentum~\cite{Khachatryan:2014rra}. However,
it is important to stress that the contact operator approximation
adopted in~\cite{Khachatryan:2014rra} only holds if the mediator is
heavy and can be integrated out~\cite{LopezHonorez:2012kv}. If the
mediator is light and contributes to resonant DM production (as in the
minimal Higgs portal model discussed herein), the contact
approximation fails and the mono-jet bounds do not apply. Future LHC14
mono-$X$ searches will also probe vertex operators for which the
mediator between dark matter and quarks is
heavy~\cite{Carpenter:2013xra,Abdallah:2015wta}, and therefore cannot
constrain the Higgs portal model discussed in this paper.

\subsection{Constraints from cosmology}

Cosmological observations further constrain the model. The earliest
observationally verified landmarks -- big bang nucleosynthesis (BBN)
and the cosmic microwave background (CMB) decoupling epoch -- have
become the de facto worldwide standard for probing theoretical
scenarios beyond the SM containing new light species. It is
advantageous to normalize the extra contribution to the SM energy
density to that of an ``equivalent'' neutrino species. The number of
``equivalent'' light neutrino species,
\begin{equation}
N_{\rm eff} = \frac{\rho_{\rm R} - \rho_\gamma}{\rho_{\nu_L}} \,,
\end{equation}
quantifies the total ``dark'' relativistic energy density (including
the three left-handed SM neutrinos) in units of the density of a
single Weyl neutrino 
\begin{equation}
\rho_{\nu_L} = \frac{7 \pi^2}{120} \ \left(\frac{4}{11} \right)^{4/3} T_\gamma^4, 
\end{equation}
where $\rho_\gamma$ is the energy density of photons (which by today
have redshifted to become the CMB photons at a temperature of about
$T_\gamma^{\rm today} \simeq 2.7~{\rm K}$)~\cite{Steigman:1977kc}.

Recent results reported by the Planck Collaboration~\cite{Ade:2015xua}
have strongly constrained the the presence of an excess $\Delta N_{\rm
  eff}$ above SM expectation: $N_{\rm eff}^{\rm SM} = 3.046$~\cite{Mangano:2005cc}. Specifically, the 68\% C.L. constraints on
$N_{\rm eff}$ from Planck TT, TE, and EE spectra, when combined with
polarization maps (lowP) and baryon acoustic oscillation (BAO)
measurements are~\cite{Ade:2015xua}:
\begin{equation*}
N_{\rm eff} = \left\{\begin{array}{cl} 3.13 \pm 0.32 & {\rm Planck TT + low
    P}, \\
3.15 \pm 0.23 & {\rm Planck TT + lowP + BAO} ,\\
2.99 \pm 0.20 & {\rm Planck TT, TE, EE+lowP},\\
3.04 \pm 0.18 & {\rm Planck TT,TE,EE+ lowP + BAO}. \end{array}
\right. 
\end{equation*}  
The joint CMB+BBN predictions on $N_{\rm eff}$ provide comparable
constraints.  The 95\% C.L. preferred range on $N_{\rm eff}$ when
combining Planck data (TT, TE, EE+lowP) with the helium abundance
estimated in~\cite{Aver:2013wba} is $N_{\rm eff} = 2.99 \pm 0.39$,
whereas the combination of Planck data with the deuterium abundance
measured in~\cite{Cooke:2013cba} yields $N_{\rm eff} = 2.91 \pm
0.37$~\cite{Ade:2015xua}. (See also~\cite{Nollett:2014lwa}.) In
summary, one fully thermalize neutrino, $\Delta N_{\rm eff} \simeq 1$,
is excluded at over 3$\sigma$. Models predicting fractional changes of
$\Delta N_{\rm eff} \approx 0.39$ are marginally consistent with data,
saturating the $1\sigma$ upper limit. Models predicting, $\Delta
N_{\rm eff} \approx 0.57$, are ruled out at about $2\sigma$.

As noted in~\cite{Weinberg:2013kea} the Goldstone boson $\alpha'$ is a
natural candidate for an imposter equivalent neutrino. The
contribution of $\alpha'$ to $N_{\rm eff}$ is $\Delta N_{\rm eff} =
\rho_{\alpha'}/\rho_{\nu}$.  Thus, taking into account the isentropic
heating of the rest of the plasma between the decoupling temperatures,
$T_{\alpha'}^{\rm dec}$ and $T_{\nu}^{\rm dec}$, we obtain
\begin{equation}
\Delta N_{\rm eff} = \frac{4}{7} \left( \frac{g (T^{\rm dec}_{\nu})}{g (T^{\rm dec}_{\alpha'})} \right)^{4/3} \,,
\label{deltaNeff}
\end{equation}
where $g(T)$ is the effective number of interacting (thermally
coupled) relativistic degrees of freedom at temperature $T$; for
example, $g(T_{\nu}^{\rm dec}) = 43/4$.\footnote{
  If relativistic particles are present that have decoupled from the
  photons, it is necessary to distinguish between two kinds of $g$:
  $g_\rho$, which is associated with the total energy density, and
  $g_s$, which is associated with the total entropy density.  For our
  calculations we use $g = g_\rho = g_s$.}  For the particle content
of the SM, there is a maximum of $g(T_{\alpha'}^{\rm dec}) =
427/4$ (with $T_{\alpha'}^{\rm dec} > m_t$). This corresponds 
to a minimum value of $\Delta N_{\rm eff} = 0.027$, which is
consistent with cosmological observations. However, a fully
thermalized 
$\alpha'$, {\it i.e.} $T^{\rm dec}_{\nu} = T^{\rm dec}_{\alpha'}$ is
excluded at 90\% C.L..  Note that if $\alpha'$ goes out
of thermal equilibrium while the temperature is just above the muon
mass 
\begin{equation}
\Delta N_{\rm eff}
= (4/7) (43/57)^{4/3} = 0.39 \, .
\end{equation}
This corresponds to a number of equivalent light neutrino species that
is consistent at the $1\sigma$ level with current data. 

The $\alpha'$ decouples from the plasma when its mean free path becomes
greater than the Hubble radius at that time.  The $\alpha'$ collision
rate with any fermion species of mass $m_f$ at or below $T$ is of
order~\cite{Weinberg:2013kea}
\begin{equation} 
\Gamma (T) \sim \frac{\lambda_3^2 m_f^2 T^7}{m_{h_1}^4 m_{h_2}^4} \, , 
\end{equation}
whereas the expansion rate of the universe is of order 
\begin{equation}
H(T) \approx \frac{T^2}{M_{\rm Pl}} \, .
\end{equation} 
We equate these two rates to obtain 
\begin{equation}
T_{\alpha'}^{\rm dec} \approx  \left(\frac{m_{h_1}^2 \,
    m_{h_2}^2}{\lambda_3 \ m_f \ M_{\rm Pl}} \right)^{1/5} \, .
\end{equation}
Now, taking  $m_f = T =  m_\mu$ we obtain
\begin{equation}
m_h \approx \frac{\left(\lambda_3^2 \, m_\mu^7 \, M_{\rm
      Pl}\right)^{1/4}}{m_H^4} \, .
\label{emeh}
\end{equation}
Substituting the conservative value $\lambda_3 = 5 \times 10^{-3}$ in
(\ref{emeh}) we have
$m_h \approx 500~{\rm MeV}$. Note that if the $\alpha'$ goes out of equilibrium when the
only massive SM particles left are $e^+ e^-$ pairs, 
$\Delta N_{\rm eff} = 0.57$. In such a case the value of $m_h$ would
have to be less than given by  (\ref{emeh}) by a
factor between $(m_e/m_\mu)^{1/2}$ and $(m_e/m_\mu)^{7/4}$~\cite{Weinberg:2013kea}. This sets a
lower limit on the mass of the hidden scalar: $m_h \approx 35~{\rm MeV}$.

\section{RG Evolution Equations}
\label{sec4}

One-loop corrections to (\ref{higgsV}) can be implemented by making
$\lambda_1$, $\lambda_2$, and $\lambda_3$ energy dependent
quantities. The positivity conditions of (\ref{VernonGabePaul}) then
must be satisfied at all energies.

A straightforward calculation leads to the RG equations for the five
parameters in the scalar potential
\begin{eqnarray} \label{RG}
 \frac{d \mu_1^2}{dt} &=&
\frac{\mu_1^2}{16\pi ^2}\left( 12\lambda _1 +6Y_t^2+2\frac{\mu_2
    ^2}{\mu_1^2}\lambda _3
  -\frac{9}{2}g_2^2-\frac{3}{2}g_Y^2\right)\,
, \nonumber \\ 
\frac{d \mu_2 ^2}{dt} &=& \frac{\mu_2^2}{16\pi
   ^2}\left( 8\lambda _2 +4\frac{\mu_1^2}{\mu_2^2}\lambda _3 +
   4f^2\right)\, , \nonumber \\  
\frac{d\lambda_1}{dt} & = &
 \frac{1}{16\pi ^2}\left( 24\lambda _1^2+\lambda _3^2
-6Y_t^4 +\frac{9}{8}g_2^4 +\frac{3}{8}g_Y^4 \right.
\nonumber \\ & + &  \left. \frac{3}{4}g_2^2g_Y^2 +
 12\lambda _1 Y_t^2 
 -   9\lambda _1 g_2^2-3\lambda _1 g_Y^2
	\right)\, ,\\ 
 \frac{d \lambda _2}{dt} &=&  \frac{1}{8\pi ^2}\left( 10\lambda
   _2^2+\lambda _3^2 - \frac{1}{4} f^4 + 4 \lambda_2 f^2
 \right)\, , \nonumber \\ 
\frac{d \lambda _3}{dt} &=&  \frac{\lambda _3}{8\pi
  ^2}\left( 6\lambda _1+4\lambda _2+2\lambda
  _3+3Y_t^2-\frac{9}{4}g_2^2    \right. \nonumber \\
& - & \left. \frac{3}{4}g_Y^2 + 2 f^2 \right) \nonumber ,
\end{eqnarray}
where $t = \ln Q$ and $Y_t$ is the top Yukawa coupling, with
\begin{equation}\label{RGE_yuk_top}
\frac{dY_t}{dt} = \frac{Y_t}{16\pi ^2}\left(
  \frac{9}{2}Y_t^2-8g_3^2-\frac{9}{4}g_2^2-\frac{17}{12}g_Y^2 \right) \,,
\end{equation}
and $Y_t^{(0)} = \sqrt{2} \, m_t/\langle \phi \rangle$ (see
Appendix~\ref{AB} for details). 
The RG running of the gauge couplings follow the standard form
\begin{eqnarray}
\frac{dg_3}{dt} &=& \frac{g_3^3}{16\pi ^2}\left[ -11+\frac{4}{3}n_g
\right] = - \frac{7}{16} \, \frac{ g_3^3}{\pi ^2} \, , \nonumber \\
\frac{dg_2}{dt} &=& \frac{g_2^3}{16\pi ^2}\left[
  -\frac{22}{3}+\frac{4}{3}n_g+\frac{1}{6}\right] = -\frac{19}{96}
\,\frac{g_2^3}{\pi ^2} \,,  \nonumber \\
\label{RGg} \frac{dg_Y}{dt}  &=& \frac{1}{16\pi ^2}\left[\frac{41}{6}
  \ g_Y^3 \right]\, , \\ 
\nonumber 
\end{eqnarray}
where $n_g =3$ is the number of generations~\cite{Arason:1991ic}. 
Finally, the running of $f$ is driven by~\cite{Iso:2009ss}
\begin{equation}
\label{RGf}\dfrac{df}{dt} = \dfrac{f^{3}}{4\pi^{2}} \, .
\end{equation}

\section{Vacuum Stability Constraints}
\label{sec5}

We now proceed to study the vacuum stability of the model through
numerical integration of Eqs.~\er{RG}, \er{RGE_yuk_top}, \er{RGg} and
\er{RGf}. To ensure perturbativity of $f$ between the TeV scale and
the Planck scale we find from (\ref{RGf-sol}), 
\begin{equation}
f=\left(\dfrac{1}{f_{0}^{2}} - \dfrac{(t-t_{0})}{2
    \pi^{2}}\right)^{-1/2} \,,
\label{RGf-sol}
\end{equation}
yielding $f_0 < 0.7$. For normalization, we set $t = \ln(Q/125~{\rm
  GeV})$ and $t_{\rm max} = \ln(\Lambda/125~{\rm GeV})$.  Now, using
the SM relation $m_H^2 = - 2 \mu^2$, with $m_H \simeq 125~{\rm GeV}$,
and setting $\langle \phi \rangle^2 = 246~{\rm GeV}$ at the same
energy scale $Q = 125~{\rm GeV}$ we fix the initial conditions for the
parameters $\mu$ and $\lambda$. Throughout we take the top Yukawa
coupling renormalized at the top pole mass~\cite{Casas:1994us}.

\subsection{Light scalar singlet}

 We integrate the RG equations from $m_{h_2} = m_H$ and impose the initial conditions for $\lambda_{1,2,3}$ by putting the observed values into \er{12}
\begin{equation}
\left. \langle \phi_{\rm SM} \rangle^2 \right|_{Q = m_{h_2}} =
\left. \langle \phi \rangle ^2 \right|_{Q = m_{h_2}} \,,\quad m_{h_2} = m_H\,.
\end{equation}
The other quantities in \er{12} $m_{h_1} = m_h, \xt$ and $\br r \ke$
remain free parameters.  It is easily seen through numerical
integration of \er{RG}, \er{RGE_yuk_top}, \er{RGg} and \er{RGf}, that
there are stable vacua up to the Planck scale. However, for those
stable vacua, the required values of $\theta$ and $m_h$ are excluded
at 90\% C.L.

As an illustration, we note that there is a stable solution for
$\langle r \rangle = 2.8~{\rm GeV}$ and $m_h = 0.3~{\rm GeV}$, which
corresponds $\theta = 0.01$. As can be seen in Fig.~\ref{rg-w_f2},
this region of the parameter space is excluded at 90\% C.L.  Actually,
for $m_h = 0.3~{\rm GeV}$, it can be shown that the mixing angle is
bounded from below: $\theta > 0.004$. The argument is as follows. The
Yukawa coupling $f$ of the Majorana fermion does not suppress the
growth of $\lambda_2$, but does exactly the opposite. This is due to
the smallness of $f$ and therefore $f^4 < 16 \lambda_2^2 f^2$ in $d
\xl_2/d t$. As a result, we can simply set $f = 0$. The RG equation of
$\xl_2$ then implies a constraint on its boundary value: $\lambda_2
\at {Q = m_H} < 0.2$ or it blows up before reaching the Planck
scale. For $\lambda_2\at{Q = m_H} = 0.2$, we need $\lambda_3\at{Q =
  m_H} < -0.28$ to have $\lambda_1$ always positive. We note that a
positive $\lambda_3$ only makes $\xl_1$ grows slower and does not help
the situation. A smaller $\lambda_2\at{Q = m_H}$ only slows down the
growth of $\lambda_3$ and does not improve the stability. In other
words, the maximum of $\lambda_3 \at{Q = m_H}$ is $-0.28$. Moreover,
from (\ref{12}) we see that the mixing angle decreases monotonically
when either $\lambda_3\at{Q = m_H}$ (when it is negative) or
$\lambda_2\at{Q = m_H}$ increases. So we reach a minimum angle when
$\lambda_2\at{Q = m_H} = 0.2$ and $\lambda_3 \at{Q = m_H} = -0.28$,
which gives $\theta = 0.004$. Such a value is excluded at the 90\%
C.L.

Next, we show that for $m_h > 0.3~{\rm GeV}$,  the required mixing
angle for a stable vacuum up to the Planck scale is $\theta > 0.004$. To this
end, we rewrite (\ref{12})  as
\begin{eqnarray}
\lambda_2 & = & \frac{m_H^2}{4y^2}  2 x^2 + \frac{m_h^2}{4y^2} ( 2 - 2
x^2)    \,,  \label{poronguita} \\
\lambda_3 & = &  2x \frac{m_h^2 - m_H^2}{2 \langle \phi \rangle y }\,,
\label{poronga}
\end{eqnarray}
where we have taken $x = \sin \theta$ and $y = \langle r \rangle$. Now,
since $m_h < m_H$ by  increasing $m_h$ we decrease $|m_h^2 - m_H^2|$
and therefore from (\ref{poronga}) we see that $x/y$
increases. Consequently, the term 
\begin{equation}
\frac{m_H^2}{4y^2} 2 x^2 - \frac{m_h^2}{4y^2} 2x^2
\end{equation}
in (\ref{poronguita}) increases (because it is proportional to
$x^2/y^2$) and therefore the other term $\propto 1/y^2$ decreases. In
other words, we have both $x/y$ and $y$ rising and therefore $x
(\theta)$ increases with increasing $m_h$.

\begin{figure}[tbp]
\postscript{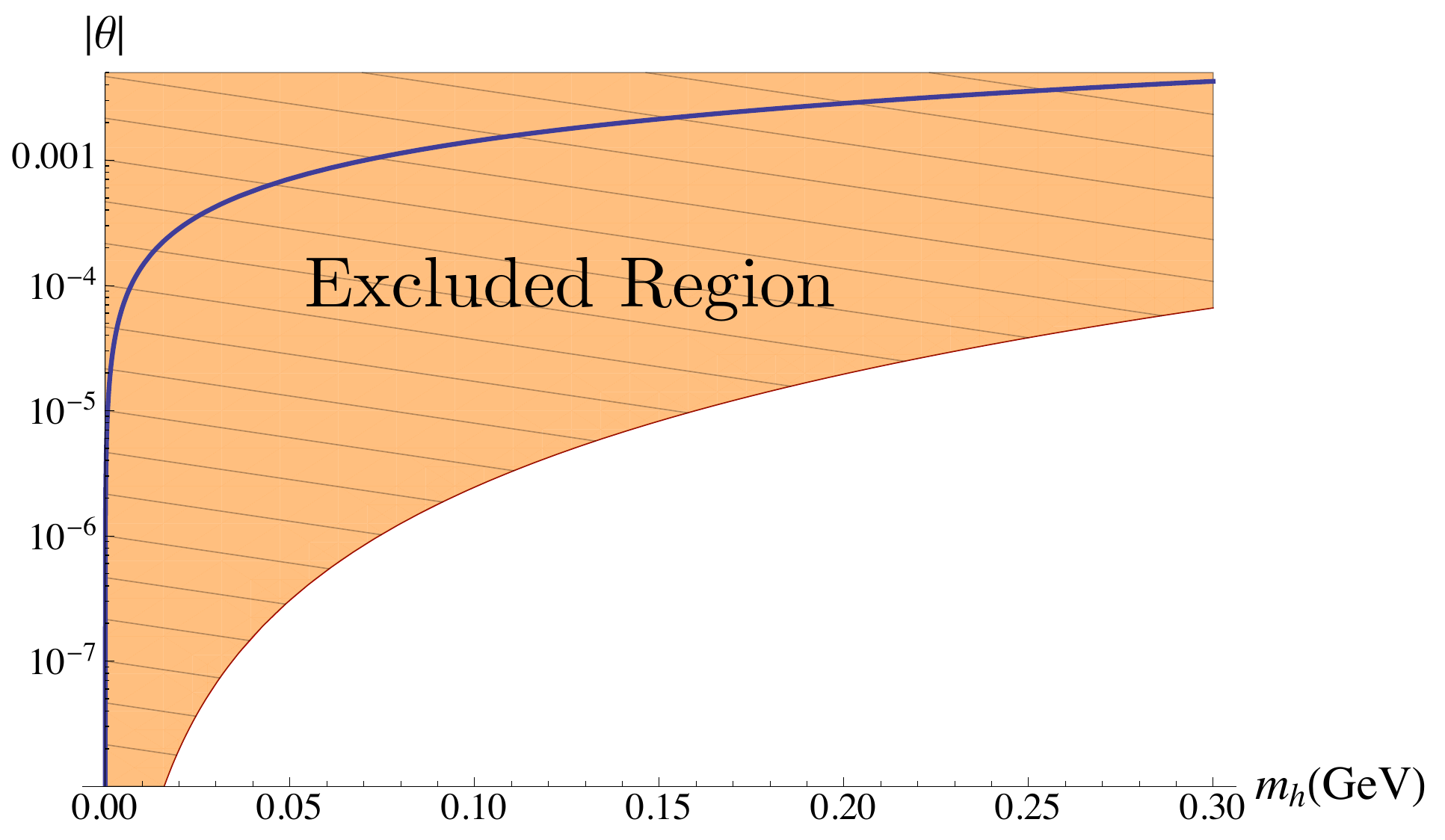}{0.99}
\caption{Comparison of vacuum stability requirements in the
  ($\theta,m_h$) plane (blue curve) with the upper limit set by SN1987A
  observations.}
\label{rg-w_f4}
\end{figure}

It should be noted that a theoretical lower limit on the mass of the
hidden scalar can be obtained by generalizing the
Weinberg-Linde~\cite{Weinberg:1976pe,Linde:1975gx} bound (see
also~\cite{Sher:1993mf}). Herein instead we have used experimental
data to determine such a lower limit.  For $m_h < 0.3~{\rm GeV}$, the
previously derived lower bound on $\theta$ can be relaxed. However,
for $m_h = 35~{\rm MeV}$, we cannot reduce the mixing angle to a level
consistent with searches for the rare flavor-changing neutral-current
decay $K^+ \to \pi^+ + E {\!\!\!\!/}_ T$ without sacrificing vacuum
stability, {\it i.e.} $\lambda_3 \sim 1$ is required to obtain $\theta
\alt 10^{-4}$. Moreover, the upper limit set by SN1987A observations
excludes values of $m_h < 35~{\rm MeV}$, for $\lambda_2 \alt 0.2$. As
an illustration, in Fig.~\ref{rg-w_f4} we show a comparison between
the $\theta$ behavior imposed by vacuum stability and the upper limit
on the mixing angle derived from (\ref{sn1987}), fixing the quartic
coupling of the hidden scalar to the fiducial value that saturates the
condition of vacuum stability, {\it i.e.} $\lambda_2 = 0.2$.

We conclude that, for $m_{h_2} = m_H$,  there are no stable solutions up to the Planck scale in
the allowed region of the parameter space.

\subsection{Heavy scalar singlet}

For energies below the mass of the heavier Higgs $h_2$, the
  effective theory is (of course) the SM. In the low energy regime the
 Higgs sector is given by
\begin{equation}
\mathscr{L}_{\rm SM} \supset ({\cal D}_\mu \Phi)^\dagger \ ({\cal D}^\mu \Phi) - \mu^2 \Phi^\dagger \Phi - \lambda (\Phi^\dagger \Phi)^2 \,, 
\label{higgsSM}
\end{equation}
and the RG equations are those of SM. To obtain the matching
conditions connecting the two theories,
following~\cite{EliasMiro:2012ay} we integrate out the field $S$ to
obtain a Lagrangian of the form (\ref{higgsSM}). Identifying the quadratic
and quartic terms in the potential yields 
\begin{equation}
\mu^2 = \mu_1^2 - \mu_2^2 \ \frac{\lambda_3}{2 \lambda_2}  
\end{equation}
and
\begin{equation}
\lambda = \lambda_1 \ \left(1 - \frac{\lambda_3^2}{4 \lambda_1 \lambda_2} \right) \, ,
\end{equation}
respectively. This is consistent with the continuity of $\langle
\phi_{\rm SM}
\rangle \leftrightharpoons 
\langle \phi \rangle$; namely
\begin{equation}
\langle \phi_{\rm SM} \rangle^2 = - \left. \frac{\mu^2}{\lambda} \right|_{Q = m_{h''}} = -
\left. \frac{\mu_1^2 - \frac{\mu_2^2 \ \lambda_3}{2 \lambda_2}}{ \lambda_1
    \ 
\left(1 -\frac{\lambda_3^2}{4 \lambda_1 \lambda_2} \right)} \right|_{Q
= m_{h_2}} \, , \nonumber 
\end{equation}
or equivalently
\begin{equation}
\left. \langle \phi_{\rm SM} \rangle^2 \right|_{Q = m_{h_2}} =
\left. \langle \phi \rangle ^2 \right|_{Q = m_{h_2}} \,,
\end{equation}
with $\langle \phi \rangle$ given by (\ref{minima}). The quartic interaction between
the heavy scalar singlet and the Higgs doublet provides an essential
contribution for the stabilization the scalar field
potential~\cite{EliasMiro:2012ay}.

When we refer to the stability of (\ref{higgsV}) at some energy $Q$
(with the use of the couplings at that scale), we are assuming that
the field values are at the scale $Q$. Note that the field values are
the only functional arguments when talking about a potential like
(\ref{higgsV}), and therefore the appropriate renormalization scale
must also be at that scale. For $\lambda_3 >0$, the third condition in
(\ref{VernonGabePaul}) could potentially be violated only for field values $\langle
\phi \rangle$ around $m_{h_2}$, regardless of the renormalization scale
$Q$~\cite{EliasMiro:2012ay}.  Consequently, the  region of instability is found
to be: 
\begin{eqnarray}
 \langle r \rangle & < & m_{h_2}/\sqrt{2\lambda_2}, \nonumber \\
  Q_- & < &
\langle \phi \rangle < Q_+, \label{rolfi}\\
     Q^2_\pm & = & \left. \frac{m_{h_2}^2 \lambda_3}{8 \lambda_1 \lambda_2} \left(1
  \pm \sqrt{1 - \frac{4\lambda_1 \lambda_2}{\lambda_3^2}} \right)
\right|_{Q_*} \,, \nonumber
\end{eqnarray}
where $Q_*$ is some energy scale where the extra positivity condition
is violated; see Appendix~\ref{AC}.\footnote{Note that (\ref{rolfi})
  is where the potential can become negative. If the third condition
  in (\ref{VernonGabePaul}) is satisfied, $Q_\pm$ will be imaginary,
  which implies that the potential is always positive. So we need to
  make sure the third condition is satisfied $Q_\pm \sim m_{h_2}$ so
  that the potential can never be negative. On the other hand, we only
  need to consider the third condition in this range as for other
  $\langle \phi\rangle$, the potential is positive regardless of the
  value of $ \frac{1}{4} \lambda_3^2 - \lambda_1 \lambda_2$.} Therefore,
$Q_\pm \sim m_{h_2}$ when the extra positivity condition is saturated,
that is $\lambda_1 \lambda_2 = \lambda_3/4$. From (\ref{rolfi}) it
follows that $Q_\pm \sim m_{h_2}$ when all the $\lambda_i$ are roughly
at the same scale. If one of the $\lambda_{1,2}$ is near zero, then
$Q_+$ can be $\gg m_{h_2}$, but this region of the parameter space is
constrained by the condition $\lambda_{1,2}>0$. The stability for
field values at $m_{h_2}$ is then determined by the potential with
coupling at scale $m_{h_2}$ (instead of $Q$). Therefore, for
$\lambda_3>0$, we impose the extra positivity condition in the
vicinity of $m_{h_2}$.  Even though the potential seems unstable at $Q
\gg m_{h_2}$, it is actually stable when all the field values are at
the scale $Q$. Note that the potential with $\lambda_i(Q)$ can only be
used when the functional arguments (field values $\langle \phi
\rangle$, $\langle r \rangle$) are at the scale $Q$. On the other
hand, the instability region for $\lambda_3<0$ is given by
\begin{eqnarray}
 \langle r \rangle & > & \frac{m_{h_2}}{\sqrt{2\lambda_2}},  \nonumber \\
 c_- & < & \frac{\langle \phi \rangle}{\langle r \rangle} <c_+, \\
c_{\pm}^2 & = & \left. - \frac{\lambda_3}{2\lambda_1} \left( 1 \pm \sqrt{1
        - \frac{4\lambda_1 \lambda_2}{\lambda_3^2} }\right)
  \right|_{Q_*}\,, \nonumber
\label{condiciones2}
\end{eqnarray}
and hence is given by the ratio of $\langle \phi \rangle$ and $\langle
r \rangle$, which can be
reached even with both $\langle \phi \rangle$ and $\langle r \rangle$ being $\gg
m_{h_2}$; see Appendix~\ref{AC}. Therefore, for $\lambda_3< 0$, we
impose the extra positivity condition at all energy scales.  Note that
the asymmetry in $\lambda_3$ will carry over into an asymmetry in
$\theta$.

\begin{figure}[tbp]
\postscript{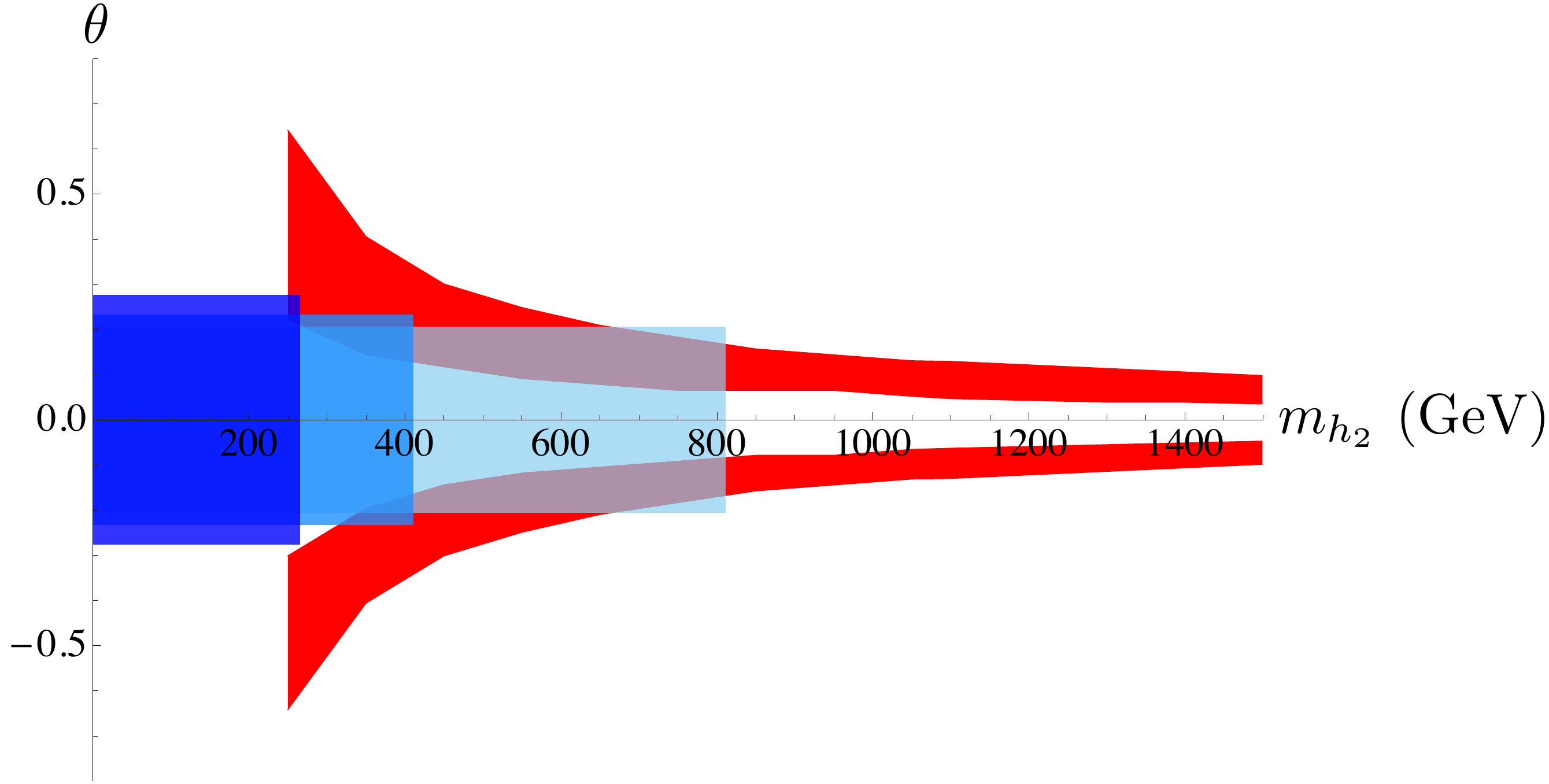}{0.99}
\caption{The red area shows the allowed parameter space in the
  $m_{h_2}$ {\it vs.} $\theta$ plane under the vacuum stability
  constraint of Eq.~(\ref{VernonGabePaul}), with $\Lambda =
  10^{19}~{\rm GeV}$. The blue areas indicate the regions of the
  parameter space that are not excluded by direct DM searches for $f_0
  = 0.4, \, 0.5, \, 0.7$, from light to dark shading. The perturbative
  upper bound is defined by $\lambda_i < 2\pi$.}
\label{rg-w_f5}
\end{figure}

\begin{figure}[tbp]
\postscript{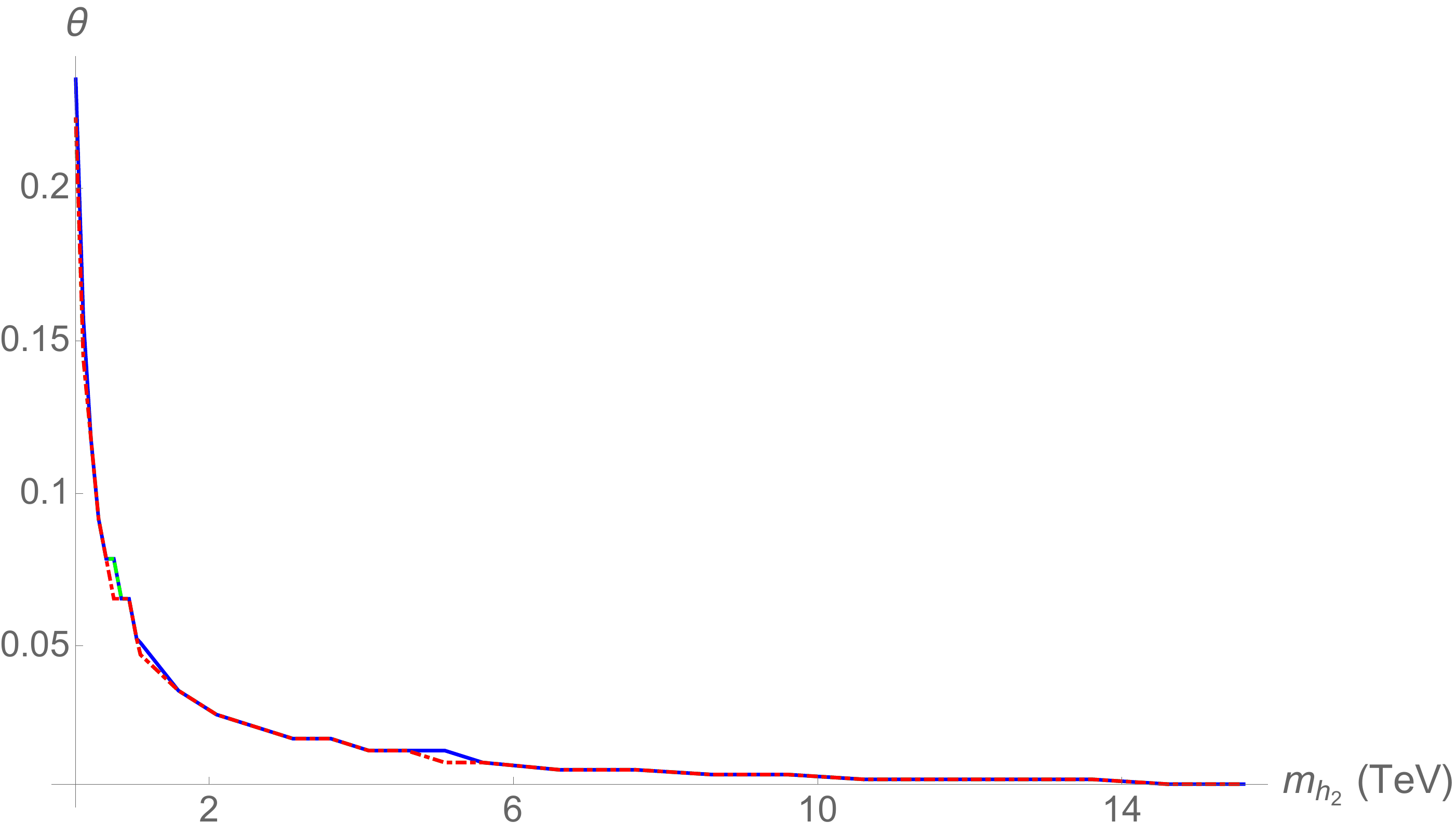}{0.99}
\caption{Comparison of three solutions of stable vacua, with identical
  initial conditions except for $f_0=0.4$ (red dashed line), 
  $f_0=0.5$ (green dot-dashed line), and $f_0 = 0.7$ (blue solid
  line).}
\label{rg-w_f6}
\end{figure}

\begin{figure}[tbp]
\postscript{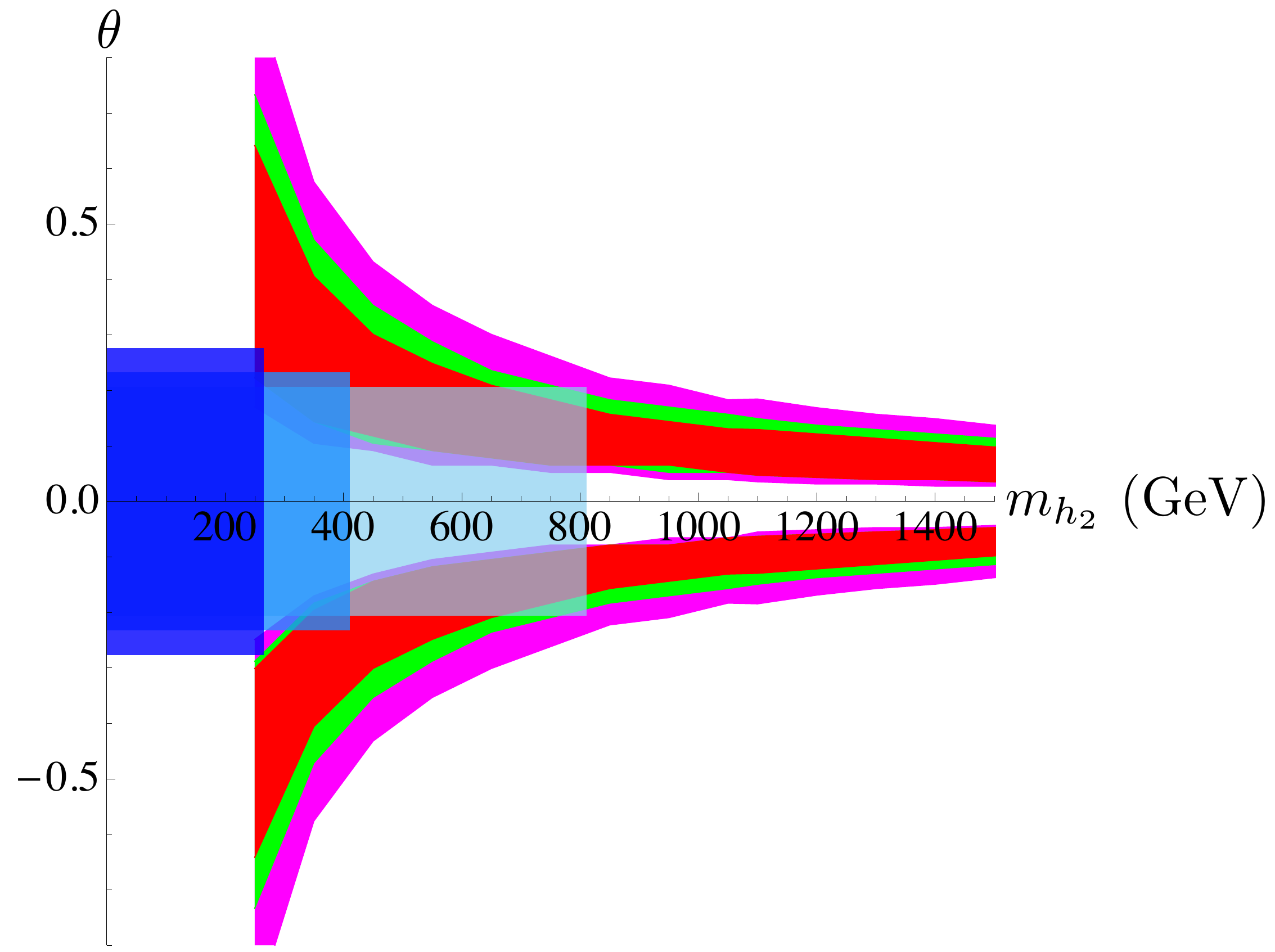}{0.99}
\caption{The allowed parameter space in the $m_{h_2}$ {\it vs.}
  $\theta$ plane under the vacuum stability constraint of
  Eq.~(\ref{VernonGabePaul}), with $\Lambda = 10^{11}~{\rm GeV}$
  (magenta), $\Lambda = 10^{15}~{\rm GeV}$ (green), and $\Lambda =
  10^{19}~{\rm GeV}$ (red).  The blue areas indicate the regions of
  the parameter space that are not excluded by direct DM searches for
  $f_0 = 0.4, \, 0.5, \, 0.7$, from light to dark shading. The
  perturbative upper bound is defined by $\lambda_i < 2\pi$.}
\label{rg-w_f7}
\end{figure}

To solve the system -- \er{RG}, \er{RGE_yuk_top}, \er{RGg} and
\er{RGf} -- we run the SM couplings from 125~GeV up to the mass scale
$m_{h_2}$ and use the matching conditions to determine $\langle
\phi_{\rm SM} \rangle$, which in turns allows one to solve
algebraically for $m_{h_1}$. In Fig.~\ref{rg-w_f5} we compare the
region of the parameter space which contains stable vacua up to the
Planck scale (red area) with the allowed (blue) bands from direct DM
searches.  From (\ref{efe}) it is straightforward to see that the
heaviest WIMP satisfying the relic density constraint, $m_w = 70~{\rm
  TeV}$, is near the unitarity limit~\cite{Griest:1989wd}. However,
one can immediately recognize in Fig.~\ref{rg-w_f1} that such a WIMP
mass exceeds the perturbativity limit, $f_0 \leq 0.7$. The maximum
WIMP mass that simultaneously satisfies the relic density constraint
in (\ref{efe}) and the $f_0$ perturbativity limit in (\ref{RGf-sol})
is $m_w = 1~{\rm TeV}$.  This maximum mass then determines the range
of the darker blue band in the horizontal axis ($m_{h_{2}}$) of
Fig.~\ref{rg-w_f5}.  The LUX upper bound on the WIMP-nucleon cross
section for elastic scattering~\cite{Akerib:2013tjd} via
(\ref{DM_bound2}) sets an upper limit on the mixing angle. The allowed
values of $\theta$ determine the range of the (blue) bands in the
vertical axis of Fig.~\ref{rg-w_f5}. The different blue bands
correspond to three fiducial values of the Majorana coupling
$f_0=0.4,\ 0.5,\ 0.7$. It is important to stress that the $f_0$
dependence of the RG running can be safely neglected; see
Fig.~\ref{rg-w_f6}.  It is also important to stress that new physics
thresholds, which may appear near the Planck mass, does not
significantly modify the region of the parameter space with stable
vacua, see Fig.~\ref{rg-w_f7}. In summary, the superposition of the
blue and red areas in Fig.~\ref{rg-w_f5} indicates the region of the
parameter space which develops a stable vacuum, satisfies the relic
density condition, and is in agreement with direct DM searches. The
interesting region of the parameter space comprises WIMP masses
$350~{\rm GeV} \alt m_w \alt 1~{\rm TeV}$.\footnote{Curiously, the
  ATLAS Collaboration has reported a $3\sigma$ excess of Higgs pair
  production $HH \to \gamma \gamma b \bar b$ for $m_{h_2} \sim
  300~{\rm GeV}$~\cite{Aad:2014yja}. See
  also~\cite{Khachatryan:2015yea}.}  The region of interest is within
reach of second generation DM direct detection experiments, such as
DEAP3600, DarkSide G2, XENONnT, and
DARWIN~\cite{Feng:2014uja,Baudis:2014naa}.

\section{Conclusions}
\label{sec6}

We have studied the vacuum stability of a minimal Higgs portal model
in which the SM particle spectrum is extended to include one complex
scalar field $S$ and one Dirac fermion field $\psi$. These new fields
are singlets under the SM gauge group and are charged under a global
$U(1)$ symmetry: $U(1)_W (\psi) =1$ and $U(1)_W(S) =2$. The
spontaneous breaking of this $U(1)$ symmetry results in a massless
Goldstone boson, a massive $CP$-even scalar, and splits the Dirac
fermion into two new mass-eigenstates $\psi_\pm$, corresponding to
Majorana fermions. The symmetry breaking yields naturally a WIMP
candidate. Fields with an even (odd) charge under the global $U(1)$
symmetry will acquire, after symmetry breaking, an even (odd) discrete
charge under a $Z_2$ discrete symmetry. While the SM particles are all
even under $Z_2$, the Majorana fermions $\psi_\pm$ are odd. In such a
set up the lightest particle with odd charge, $\psi_-$, will be
absolutely stable, and hence a plausible WIMP candidate. 

We have shown that interactions between the extended Higgs sector and
the lightest Majorana fermion which are strong enough to yield a
thermal relic abundance consistent with observation can easily
destabilize the electroweak vacuum or drive the theory into a
non-perturbative regime at an energy scale well below the Planck
mass. However, we have also unmasked a small region of the parameter
space which develops a stable vacuum (up to the Planck scale),
satisfies the relic abundance, and is in agreement with direct DM
searches. This region comprises WIMP masses $350~{\rm GeV} \alt m_w
\alt 1~{\rm TeV}$. The region of interest is  within
reach of second generation DM direct detection experiments.

Needless to say, here we have considered a minimal model to ensure
that bounding the parameter space remains tractable. However, our
extension of the dark sector enlarges the parameter space sufficiently
to contain stable vacua up to the Planck scale.

\acknowledgments{We thank Wai-Yee Keung and Neal Weiner for valuable discussions. LAA
  is supported by U.S. National Science Foundation (NSF) CAREER Award
  PHY1053663 and by the National Aeronautics and Space Administration
  (NASA) Grant No. NNX13AH52G; he thanks the Center for Cosmology and
  Particle Physics at New York University for its hospitality. VB is
  supported by the U. S. Department of Energy (DoE) Grant No. DE-
  FG-02- 95ER40896. HG is supported by NSF Grant No. PHY-1314774.  XH
  is supported by the MOST Grant 103-2811-M-003-024. DM is supported
  by DoE Grant No. DE-SC0010504. TJW is supported by DoE Grant
  No. DE-SC0011981 and the Simons Foundation Grant No. 306329.}

\onecolumngrid

\appendix

\section{}
\label{AA}

To determine the value of $G_{q}/m_q$ we look back at (\ref{eq:fifty})
along with the SM Yukawa interaction term, which involves the mixing
of both scalar fields, $h_1$ and $h_2$. For interactions of WIMPs with
SM quarks, the relevant terms are 
\begin{equation}
\mathscr{L}  =  \frac{m_q \cos
  \theta}{\langle \phi \rangle} h_{1,2} \bar \psi_q \psi_q - \frac{m_q
  \sin \theta}{\langle \phi \rangle} h_{2,1}
\bar \psi_q \psi_q + \dots 
 +  \frac{f \sin \theta}{2} h_{1,2} \bar \psi_- \psi_-
+ \frac{f \cos \theta}{2} h_{2,1} \bar \psi_- \psi_-.  
\end{equation} 
The scattering of
a $w$ particle off a quark then gives \beqa \CM &=&  i \frac{f m_q \sin
  \theta \cos \theta}{\langle \phi \rangle}  \ \bar u_q (p') \ u_q(p)  \lpa \frac{1}{t-m_{h_{1,2}}^2}
- \frac{1}{t-m_{h_{2,1}}^2} \rpa  \ \bar u(k') \ u(k) \nonumber
\\
&\approx& i \frac{f m_q \lambda_3 \langle r \rangle }{m_{h_1}^2 m_{h_2}^2} \bar u_q (p')
u_q(p) \bar u(k') u(k) .  \eeqa This leads to the identification
of the effective coupling \beq \frac{2 G_{q}}{\sqrt{2}} = \frac{m_q f
  \lambda_3 \langle r \rangle }{m_{h_1}^2 m_{h_2}^2} \Rightarrow \frac{G_{q}}{m_q} =
\frac{\sqrt{2} f \lambda_3 \langle r \rangle }{2 \ m_{h_1}^2 m_{h_2}^2}.
\label{eq:effectiveCoup}
\end{equation}

\section{}
\label{AB}

To establish the one-loop RG equations for the parameters of the scalar
potential, we first compute the one-loop
effective potential and then impose its independence from the
renormalisation scale. To one-loop level, the scalar potential is
given by $V=V^{(0)}+\Delta V^{(1)}$,
where $V^{(0)}$ is the tree-level potential and $\Delta V^{(1)}$
indicates the one-loop correction to it. To compute the latter it is useful to re-write the tree-level
potential (\ref{higgsV}) in terms of the real scalar fields: 
\begin{equation}
\Phi =
\frac{1}{\sqrt{2}} \left( \begin{array}{c}  \varphi_1 + i \varphi_2
    \\ \varphi_3 + i
    \varphi_4 \end{array} \right) \quad \quad {\rm and}  \quad \quad S = \frac{1}{\sqrt{2}} \left(\varkappa_1 +
    i \varkappa_2 \right) \, .
\end{equation}
The particular combination of fields relevant for the calculation
are $\varphi^2 =
\varphi_1^2 + \varphi_2^2 + \varphi_3^2 + \varphi_4^2$ and
$\varkappa^2 = \varkappa_1^2 + \varkappa_2^2$; hence (\ref{higgsV}) can be
rewritten as
\begin{equation}\label{tree}
V^{(0)}(\varphi,r )= \frac{1}{2}\mu_1^2 \varphi^2 +\frac{1}{2}\mu_2 ^2 \varkappa^2 +\frac{1}{4}\lambda _1 \varphi ^4 +\frac{1}{4}\lambda _2 \varkappa^4 +
			\frac{1}{4}\lambda _3 \varphi ^2 \varkappa^2\, .
\end{equation}
In the Landau gauge the one-loop correction to the tree-level potential (\ref{tree}) reads:
\begin{equation}\label{1-loop}
\Delta V^{(1)}(\varphi ,\varkappa )=\frac{1}{64\pi ^2}\sum
_i(-1)^{2s_i}(2s_i+1)M^4_i(\varphi ^2, \varkappa^2)\left[ \ln{\frac{M^2_i(\varphi ^2,\varkappa ^2)}{Q^2}-c_i}\right]\, ,
\end{equation}
where $c_i$ are constants that depend on the renormalisation scheme.
For the $\overline{\rm MS}$ scheme, we have  $c_i=5/6$ for vectors, and $c_i=3/2$
for scalars and fermions. Next, we expand
(\ref{1-loop}) and we just keep the contributions from the scalar
fields,  the top-quark, the gauge bosons, and the Majorana fermions, 
\begin{eqnarray}
\Delta V^{(1)}&=&\frac{1}{64\pi ^2}\left\{ 3 {\cal G}_1^2\left[
    \ln{\frac{{\cal G}_1}{Q ^2}-\frac{3}{2}}\right] + {\cal
    G}_2^2\left[ \ln{\frac{{\cal G}_2}{Q ^2}-\frac{3}{2}}\right]
		+ {\rm Tr}\left({\cal  H}^2\left[ \ln{\frac{{\cal
                          H}}{Q^2}-\frac{3}{2}}\right]\right)
                -12 \ T_\varphi^2\left[ \ln{\frac{T_\varphi}{Q ^2} - \frac{3}{2}}\right] 
                  \right. \nonumber \\
               & + & \left. 3{\rm Tr}\left(  M_\varphi^2\left[
              \ln{\frac{M_\varphi}{Q ^2}-\frac{5}{6}}\right]\right)
          - 4 W_\varkappa^2\left[ \ln{\frac{W_\varkappa}{Q^2}-\frac{3}{2}}\right] \right\}\, ,
\end{eqnarray}
where  (in a self-explanatory notation) the field-dependent squared masses are,
\begin{eqnarray}
{\cal G}_1 (\varphi, \varkappa )&=& \mu_1^2+\lambda _1 \varphi
^2+\frac{\lambda _3}{2} \varkappa ^2\, ,\\ 
{\cal G}_2 (\varphi ,\varkappa )&=& \mu_2 ^2+\lambda _2 \varkappa ^2+\frac{\lambda
  _3}{2} \varphi ^2\, ,\\ \label{Higgs-field-dep}
{\cal H} (\varphi ,\varkappa )&=& \left( \begin{array}{cc} \mu_1^2+3\lambda
    _1 \varphi ^2+\frac{\lambda _3}{2} \varkappa ^2 & \lambda _3 \varphi \varkappa\\ 
				\lambda _3 \varphi \varkappa & \mu_2 ^2
                                +3\lambda _2 \varkappa^2+\frac{\lambda
                                  _3}{2} \varphi ^2\end{array}\right)\, ,\\ \label{Top-field-dep}
T_\varphi (\varphi)&=& \frac{1}{2} \left(Y_t \varphi \right)^2\, ,\\ \label{RH-N-field-dep}
M_\varphi(\varphi) &=& \frac{1}{4}\left(
		\begin{array}{cc}
		g_Y^{\phantom{o}2}\varphi ^2 & -g_2g_Y\varphi ^2 \\
		-g_2g_Y\varphi ^2 & g_2^2\varphi ^2 \\
		\end{array} \right) \, ,\\ \label{Gauge_bosonos-dep}
W_\varkappa (\varkappa )&=&\frac{1}{4} \left(f \varkappa \right)^2\, .
\end{eqnarray}
We define the beta functions $\beta _i$ ($i=1\dots 3$) for the quartic
couplings, the gamma functions $\gamma _{\mu_1,\mu_2}$ for the scalar
masses, and the scalar anomalous dimensions $\gamma
_{\varphi,\varkappa}$  according to: $d\lambda _i/dt = \beta _i$,
$d\mu_1^2/dt = \gamma _{\mu_1} \mu_1^2$, $d\mu_2 ^2/dt = \gamma
_{\mu_2} \mu_2 ^2$, $d\varphi ^2/dt = 2\gamma _\varphi \varphi^2$, and
$d\varkappa ^2/dt = 2\gamma _\varkappa \varkappa^2$. We then 
extract the RG equations for the parameters of the scalar potential by
forcing the first derivative of the effective potential with respect
to the scale $t$ to vanish
\begin{equation}\label{effective-P}
\frac{d}{dt}V^{(1)} \equiv \frac{d}{dt}(V^{(0)}+\Delta V^{(1)})\equiv 0\, ,
\end{equation}
keeping only the one-loop terms. After a bit of
algebra~(\ref{effective-P}) leads to the following equations: 
\begin{eqnarray} \label{RG2}
\frac{\mu_1^2 \varphi^2}{2} \left[ \gamma _{\mu_1} +2\gamma _\varphi
  -\frac{1}{16\pi ^2}\left( 12\lambda _1+2\frac{\mu_2
      ^2}{\mu_1^2}\lambda _3\right) 
\right] &=& 0\, , \nonumber \\ 
\frac{\mu_2 ^2 \varkappa ^2}{2} \left[\gamma _{\mu_2} +2\gamma _\varkappa
  -\frac{1}{16\pi ^2}\left( 8\lambda
    _2+4\frac{\mu_1^2}{\mu_2^2}\lambda _3\right) \right] &=& 0\, ,
\nonumber \\ 
\frac{\varphi ^4}{4}\left[\beta _1 +4\lambda _1\gamma _\varphi -\frac{1}{16\pi ^2}\left( 24\lambda _1^2+\lambda _3^2
	-6Y_t^4
	+\frac{9}{8}g_2^4+\frac{3}{8}g_Y^4+\frac{3}{4}g_2^2g_Y^2
      \right)\right] &=&0\, , \\ 
\frac{\varkappa ^4}{4}\left[\beta _2 +4\lambda _2\gamma _\varkappa -\frac{1}{8\pi ^2}\left( 10\lambda _2^2+\lambda _3^2
	-\frac{1}{4} f^4 \right)\right] &=&0\, , \nonumber \\ 
\frac{\varphi ^2 \varkappa ^2}{4} \left[\beta _3 +2\lambda _3(\gamma _\varphi
  +\gamma _\varkappa)-\frac{1}{8\pi ^2}\left( 6\lambda _1\lambda _3+4\lambda
    _2\lambda _3 +2\lambda _3^2 \right)\right] &=&0\, . \nonumber
\end{eqnarray}
Requiring that each term between squared brackets vanishes, we arrive at
the RG equations for the parameters of the scalar potential. Namely,
substituting the explicit expression of the scalar anomalous
dimensions~\cite{Sher:1988mj}
\begin{equation}
\gamma _\varphi = -\frac{1}{16 \pi ^2} \left( 3Y_t^2
  -\frac{9}{4}g_2^2-\frac{3}{4}g_Y^2 \right) \quad \quad {\rm and}
\quad \quad 
\gamma _\varkappa  = -\frac{1}{8 \pi ^2}   \, f^2 \, ,
\end{equation}
into (\ref{RG2})  we obtain (\ref{RG}).

\section{}
\label{AC}

To explore the impact of the complex singlet scalar on the stability
of the Higgs sector we follow~\cite{EliasMiro:2012ay} and consider
a tree level scalar potential of the form 
\begin{equation}
V \left(\Phi, S \right)  =   \lambda_1  \left(\Phi^\dagger \Phi - \frac
{\langle \phi \rangle^{2}}
2\right)^{2} + \lambda_2 \left(S^\dagger S -\frac {\langle r \rangle^{2}}
2 \right)^{2} 
 +  \lambda_3 \left(\Phi^\dagger \Phi -\frac {\langle \phi \rangle^{2}}
2\right)\left(S^\dagger S -\frac {\langle r \rangle^{2}} 2\right) \, .
\label{higgsVnew}
\end{equation}
For $\xl_3>0$, the third term can only be negative when either one of
the factors is negative. The parameter space for $\Phi^\dagger \Phi<
{\langle \phi \rangle^{2}}/2$ is, in principle,  described by the
effective potential of the SM (with one Higgs).  So herein we only
consider $S^\dagger S< {\langle r \rangle^{2}}/2$. As argued
in~\cite{EliasMiro:2012ay}, the most dangerous region of the field configuration is given by $S =0$.\footnote{The instability region is defined by both relations $Q_- <
\Phi < Q_+$ and $\lambda_1 \lambda_2 < (2 \lambda_{3})^{-2}$, with the
couplings evaluated at the scale $\Phi$. The second relation is  more
likely to be satisfied at a high energy scale, and therefore $|\Phi|= Q_+$ 
is the most dangerous region of the field configuration to reach the
instability region, {\it i.e.} $V(\Phi, S)=0$.} In this region, we have
\begin{equation}
V \left(\Phi, 0 \right)  =   \lambda_1 (Q)  \left(\left| \Phi \right|^2
-\frac {\langle \phi \rangle^{2}}
2\right)^{2} + \lambda_2(Q) \left(\frac {\langle r \rangle^{2}}
2\right)^{2} 
 - \frac {\langle r \rangle^{2}} 2\lambda_3(Q) \left(\left| \Phi
\right|^2 -\frac {\langle \phi \rangle^{2}}
2\right) \, .
\end{equation}
The couplings are now replaced by their values at some scale $Q$. We
take $\langle \phi \rangle$ and $\langle r \rangle$ to be the physical
VEV and only the couplings $\xl_{i}$ run. This is possible in some
renormalization scheme (like taking vacuum expectation $|\Phi| =
\langle \phi \rangle, |S| = \langle r \rangle$ as one of the
renormalization conditions, which is satisfied trivially for this
particular form of potential).  Keeping only terms with $\langle r
\rangle$ (since $\langle r \rangle \gg \langle \phi \rangle$), the
condition $V=0$ can be rewritten as, \be \lambda_1 (Q) | \Phi |^4 +
\frac {\lambda_2(Q) \langle r \rangle ^{4}} 4 -\frac {\lambda_3(Q)
  \langle r \rangle^{2}} 2 \left| \Phi \right|^2 =0.  
\label{A3}
\ee Next, we
assume that $\lambda_2(Q) \langle r \rangle ^2 \sim -\mu_2(Q)^2$ is
almost unchanged under the RG flow and remains $\frac 1 2 m_{h_2}^2$ 
({\it i.e.} we assume that $\xl_i$ does not run by much). Under this
assumption (\ref{A3}) becomes 
\be \lambda_1 (Q) | \Phi |^4 -\frac
{\lambda_3(Q)m_{h_2}^2} {4 \xl_2(Q)} \left| \Phi \right|^2+ \frac {m_{h_2}^4}
{16 \lambda_2 (Q)} =0. \label{eqstability} \ee 
The solution to this equation gives Eq.~(\ref{rolfi}); the
first condition comes from $|S|^2< {\langle r \rangle^{2}}/2 \sim \frac 1 2
m_{h_2}^2$, with $\langle \phi
\rangle = \sqrt{2} |\Phi|$~\cite{EliasMiro:2012ay}.

For $\xl_3<0$, we can consider a field configuration with both
$|\Phi|\sim Q$, $|S|\sim Q$ much larger than $\langle r \rangle$. The
point is that we only need to find a configuration in which the
stability is violated. In this case, we must keep only the quartic
term and the potential becomes
\be
V \left(\Phi, S \right) =  \lambda_1 | \Phi |^4 + \lambda_2 | S |^4 +
\lambda_3 \left| \Phi 
\right|^2 \left| S \right|^2 \, .
\label{quarticp}
\ee On the one hand, following~\cite{EliasMiro:2012ay} we can
duplicate the procedure to obtain
(\ref{condiciones2}). These conditions can be satisfied and therefore
the vacuum becomes unstable. On the other hand, we can just consider
the eigenvalues of the matrix \be\bay{cc} \xl_1 & \frac 1 2 \xl_3 \\
\frac 1 2 \xl_3 & \xl_2 \eay.  \ee In fact, the second approach also
tells us why in the case of $\xl_3>0$, a potential with the form of
\eqref{quarticp}  is in fact stable. The
eigenvector with the negative eigenvalue is given by
\[\lb -\frac{-\lambda_1 +\lambda_2+\sqrt{\lambda_1^2-2 \lambda_1 \lambda_2+\lambda_2^2+\lambda_3^2}}{\lambda _3},1\rb\]
When $\xl_3^2 \ge 4 \xl_1 \xl_2$, the first component is negative. So
it requires either $|\Phi|^2$ or $|S|^2$ to be negative, which is
impossible.  As a result, for $\xl_3>0$ we need to consider a
particular field configuration to study the instability.

We now relate the two functional forms of the Higgs potential.
At the classical level (\ref{higgsV}) differs from
(\ref{higgsVnew}) by a constant; that is the vacuum energy is
shifted. In fact (\ref{higgsV}) has a negative vacuum energy $\sim -
\frac 1 4 \xl_2 (\langle r \rangle) \, \langle r \rangle^4$ (again
neglecting all $\langle \phi \rangle$ corrections) and the instability requires the
potential to be smaller than this negative vacuum energy.

At a particular scale $Q$, all the couplings $\xl_i$ in (\ref{higgsV})
can be replaced by $\xl_i(Q)$ and $\mu_{1,2}(Q)$, so that
(\ref{higgsV}) can be rewritten in the form of (\ref{higgsVnew}) with
some $\langle \phi (Q) \rangle, \langle r (Q) \rangle$ as a
combination of $\xl_{i}(Q)$ and $\mu_i(Q)$. Note that we can still
adopt our previous arguments to consider only the configuration $|S| =
0$. In this case, \be V \left(\Phi, 0 \right) = \mu_1^2(Q) \left| \Phi
\right|^2 + \lambda_1(Q) \left| \Phi \right|^4 \ee 
The condition for
stability is saturated when 
\begin{equation} 
\mu_1^2(Q) \left| \Phi \right|^2 +
\lambda_1(Q) \left| \Phi \right|^4 +\frac 1 4 \xl_2 (\langle r
\rangle) \ \langle r \rangle ^4 =0 \, .
\end{equation} 
Solving (\ref{ecuacionnueve})   we have $-\mu_1^2(\langle r \rangle)
= \frac 1 2 \xl_3(\langle r \rangle) \, \langle r \rangle^2$. Now,
assuming that all the $\xl_i$ do not run too much along the RG flow
we obtain \eqref{eqstability}.

When $\xl_{1,2}$ remains relative away from zero, (\ref{rolfi}) remains
a reasonable approximation for the scale $Q_\pm$ between which (\ie
$Q_- < \sqrt 2 |\Phi| < Q_+$) the potential can become negative. Note
that a na\"{\i}ve argument for instability using only the quartic potential
(which is usually how we get to $\xl_3^2 \ge 4 \xl_1 \xl_2$) is only
valid for $\xl_3<0$. As a result, the potential can only become
unstable in a some very particular field configuration. In this
region, however, the effective potential is not valid since the field
values are far away from the scale $Q$.

\twocolumngrid


\begin{thebibliography}{999}

\bibitem{ATLAS:2012ae}
  G.~Aad {\it et al.}  [ATLAS Collaboration],
  Phys.\ Lett.\ B {\bf 710}, 49 (2012)
  [arXiv:1202.1408 [hep-ex]].


\bibitem{Chatrchyan:2012tx} 
  S.~Chatrchyan {\it et al.}  [CMS Collaboration],
  Phys.\ Lett.\ B {\bf 710}, 26 (2012)
  [arXiv:1202.1488 [hep-ex]].






\bibitem{Aad:2013wqa} 
  G.~Aad {\it et al.}  [ATLAS Collaboration],
  Phys.\ Lett.\ B {\bf 726}, 88 (2013)
  [Phys.\ Lett.\ B {\bf 734}, 406 (2014)]
  [arXiv:1307.1427 [hep-ex]].


\bibitem{Chatrchyan:2013mxa} 
  S.~Chatrchyan {\it et al.}  [CMS Collaboration],
  Phys.\ Rev.\ D {\bf 89},  092007 (2014)
  [arXiv:1312.5353 [hep-ex]].



\bibitem{Aad:2014aba} 
  G.~Aad {\it et al.}  [ATLAS Collaboration],
  Phys.\ Rev.\ D {\bf 90},  052004 (2014)
  [arXiv:1406.3827 [hep-ex]].


\bibitem{Khachatryan:2014ira} 
  V.~Khachatryan {\it et al.}  [CMS Collaboration],
  Eur.\ Phys.\ J.\ C {\bf 74},  3076 (2014)
  [arXiv:1407.0558 [hep-ex]].



\bibitem{Lindner:1988ww} 
  M.~Lindner, M.~Sher and H.~W.~Zaglauer,
  Phys.\ Lett.\ B {\bf 228}, 139 (1989).




\bibitem{Sher:1988mj} 
  M.~Sher,
  Phys.\ Rept.\  {\bf 179}, 273 (1989).




\bibitem{Diaz:1994bv} 
  M.~A.~Diaz, T.~A.~ter Veldhuis and T.~J.~Weiler,
  Phys.\ Rev.\ Lett.\  {\bf 74}, 2876 (1995)
  [hep-ph/9408319].






\bibitem{Casas:1994qy} 
  J.~A.~Casas, J.~R.~Espinosa and M.~Quiros,
  Phys.\ Lett.\ B {\bf 342}, 171 (1995)
  [hep-ph/9409458].



\bibitem{Diaz:1995yv} 
  M.~A.~Diaz, T.~A.~ter Veldhuis and T.~J.~Weiler,
  Phys.\ Rev.\ D {\bf 54}, 5855 (1996)
  [hep-ph/9512229].



\bibitem{Casas:1996aq} 
  J.~A.~Casas, J.~R.~Espinosa and M.~Quiros,
  Phys.\ Lett.\ B {\bf 382}, 374 (1996)
  [hep-ph/9603227].


\bibitem{Isidori:2001bm} 
  G.~Isidori, G.~Ridolfi and A.~Strumia,
  Nucl.\ Phys.\ B {\bf 609}, 387 (2001)
  [hep-ph/0104016].



\bibitem{Isidori:2007vm} 
  G.~Isidori, V.~S.~Rychkov, A.~Strumia and N.~Tetradis,
  Phys.\ Rev.\ D {\bf 77}, 025034 (2008)
  [arXiv:0712.0242 [hep-ph]].


\bibitem{Hall:2009nd} 
  L.~J.~Hall and Y.~Nomura,
  JHEP {\bf 1003}, 076 (2010)
  [arXiv:0910.2235 [hep-ph]].


\bibitem{Ellis:2009tp} 
  J.~Ellis, J.~R.~Espinosa, G.~F.~Giudice, A.~Hoecker and A.~Riotto,
  Phys.\ Lett.\ B {\bf 679}, 369 (2009)
  [arXiv:0906.0954 [hep-ph]].


\bibitem{EliasMiro:2011aa} 
  J.~Elias-Miro, J.~R.~Espinosa, G.~F.~Giudice, G.~Isidori, A.~Riotto and A.~Strumia,
  Phys.\ Lett.\ B {\bf 709}, 222 (2012)
  [arXiv:1112.3022 [hep-ph]].


\bibitem{Bezrukov:2012sa} 
  F.~Bezrukov, M.~Y.~Kalmykov, B.~A.~Kniehl and M.~Shaposhnikov,
  JHEP {\bf 1210}, 140 (2012)
  [arXiv:1205.2893 [hep-ph]].

\bibitem{Degrassi:2012ry} 
  G.~Degrassi, S.~Di Vita, J.~Elias-Miro, J.~R.~Espinosa, G.~F.~Giudice, G.~Isidori and A.~Strumia,
  JHEP {\bf 1208}, 098 (2012)
  [arXiv:1205.6497 [hep-ph]].

\bibitem{Buttazzo:2013uya} 
  D.~Buttazzo, G.~Degrassi, P.~P.~Giardino, G.~F.~Giudice, F.~Sala, A.~Salvio and A.~Strumia,
  JHEP {\bf 1312}, 089 (2013)
  [arXiv:1307.3536 [hep-ph]].



\bibitem{Branchina:2013jra} 
  V.~Branchina and E.~Messina,
  Phys.\ Rev.\ Lett.\  {\bf 111}, 241801 (2013)
  [arXiv:1307.5193 [hep-ph]].



\bibitem{Branchina:2014rva} 
  V.~Branchina, E.~Messina and M.~Sher,
  Phys.\ Rev.\ D {\bf 91},  013003 (2015)
  [arXiv:1408.5302 [hep-ph]].



\bibitem{Lalak:2014qua} 
  Z.~Lalak, M.~Lewicki and P.~Olszewski,
  JHEP {\bf 1405}, 119 (2014)
  [arXiv:1402.3826 [hep-ph]].


\bibitem{Basso:2010jm} 
  L.~Basso, S.~Moretti and G.~M.~Pruna,
  Phys.\ Rev.\ D {\bf 82}, 055018 (2010)
  [arXiv:1004.3039 [hep-ph]].


\bibitem{Kadastik:2011aa} 
  M.~Kadastik, K.~Kannike, A.~Racioppi and M.~Raidal,
  JHEP {\bf 1205}, 061 (2012)
  [arXiv:1112.3647 [hep-ph]].


\bibitem{EliasMiro:2012ay} 
  J.~Elias-Miro, J.~R.~Espinosa, G.~F.~Giudice, H.~M.~Lee and A.~Strumia,
  JHEP {\bf 1206}, 031 (2012)
  [arXiv:1203.0237 [hep-ph]].



\bibitem{Cheung:2012nb} 
  C.~Cheung, M.~Papucci and K.~M.~Zurek,
  JHEP {\bf 1207}, 105 (2012)
  [arXiv:1203.5106 [hep-ph]].


\bibitem{Anchordoqui:2012fq} 
  L.~A.~Anchordoqui, I.~Antoniadis, H.~Goldberg, X.~Huang, D.~Lust, T.~R.~Taylor and B.~Vlcek,
  JHEP {\bf 1302}, 074 (2013)
  [arXiv:1208.2821 [hep-ph]].


\bibitem{Baek:2012uj} 
  S.~Baek, P.~Ko, W.~I.~Park and E.~Senaha,
  JHEP {\bf 1211}, 116 (2012)
  [arXiv:1209.4163 [hep-ph]].


\bibitem{Chao:2012mx} 
  W.~Chao, M.~Gonderinger and M.~J.~Ramsey-Musolf,
  Phys.\ Rev.\ D {\bf 86}, 113017 (2012)
  [arXiv:1210.0491 [hep-ph]].


\bibitem{Coriano:2014mpa} 
  C.~Coriano, L.~Delle Rose and C.~Marzo,
  Phys.\ Lett.\ B {\bf 738}, 13 (2014)
  [arXiv:1407.8539 [hep-ph]].



\bibitem{Altmannshofer:2014vra} 
  W.~Altmannshofer, W.~A.~Bardeen, M.~Bauer, M.~Carena and J.~D.~Lykken,
  JHEP {\bf 1501}, 032 (2015)
  [arXiv:1408.3429 [hep-ph]].

\bibitem{Falkowski:2015iwa} 
  A.~Falkowski, C.~Gross and O.~Lebedev,
  JHEP {\bf 1505}, 057 (2015)
  [arXiv:1502.01361 [hep-ph]].


\bibitem{Krog:2015cna} 
  J.~Krog and C.~T.~Hill,
  arXiv:1506.02843 [hep-ph].

\bibitem{Rose:2015fua} 
  L.~D.~Rose, C.~Marzo and A.~Urbano,
  arXiv:1506.03360 [hep-ph].




\bibitem{Feng:2010gw} 
  J.~L.~Feng,
  Ann.\ Rev.\ Astron.\ Astrophys.\  {\bf 48}, 495 (2010)
  [arXiv:1003.0904 [astro-ph.CO]].



\bibitem{Schabinger:2005ei} 
  R.~Schabinger and J.~D.~Wells,
  Phys.\ Rev.\ D {\bf 72}, 093007 (2005)
  [hep-ph/0509209].


\bibitem{Patt:2006fw} 
  B.~Patt and F.~Wilczek,
  hep-ph/0605188.


\bibitem{Barger:2007im} 
  V.~Barger, P.~Langacker, M.~McCaskey, M.~J.~Ramsey-Musolf and G.~Shaughnessy,
  Phys.\ Rev.\ D {\bf 77}, 035005 (2008)
  [arXiv:0706.4311 [hep-ph]].


\bibitem{Barger:2008jx} 
  V.~Barger, P.~Langacker, M.~McCaskey, M.~Ramsey-Musolf and G.~Shaughnessy,
  Phys.\ Rev.\ D {\bf 79}, 015018 (2009)
  [arXiv:0811.0393 [hep-ph]].

\bibitem{Krauss:1988zc} 
  L.~M.~Krauss and F.~Wilczek,
  Phys.\ Rev.\ Lett.\  {\bf 62}, 1221 (1989).



\bibitem{Weinberg:2013kea} 
  S.~Weinberg,
  Phys.\ Rev.\ Lett.\  {\bf 110},  241301 (2013)
  [arXiv:1305.1971 [astro-ph.CO]].


\bibitem{Sikivie:1982qv} 
  P.~Sikivie,
  Phys.\ Rev.\ Lett.\  {\bf 48}, 1156 (1982).


\bibitem{Vilenkin:1982ks} 
  A.~Vilenkin and A.~E.~Everett,
  Phys.\ Rev.\ Lett.\  {\bf 48}, 1867 (1982).





\bibitem{Turner:1990uz} 
  M.~S.~Turner and F.~Wilczek,
  Phys.\ Rev.\ Lett.\  {\bf 66}, 5 (1991).

\bibitem{Dvali:1991ka} 
  G.~R.~Dvali,
  Phys.\ Lett.\ B {\bf 265}, 64 (1991).


\bibitem{Hiramatsu:2012sc} 
  T.~Hiramatsu, M.~Kawasaki, K.~'i.~Saikawa and T.~Sekiguchi,
  JCAP {\bf 1301}, 001 (2013)
  [arXiv:1207.3166 [hep-ph]].




\bibitem{Ellis:2000ds} 
  J.~R.~Ellis, A.~Ferstl and K.~A.~Olive,
  Phys.\ Lett.\ B {\bf 481}, 304 (2000)
  [hep-ph/0001005].



\bibitem{Beltran:2008xg} 
  M.~Beltran, D.~Hooper, E.~W.~Kolb and Z.~C.~Krusberg,
  Phys.\ Rev.\ D {\bf 80}, 043509 (2009)
  [arXiv:0808.3384 [hep-ph]].

\bibitem{Hill:2014yxa} 
  R.~J.~Hill and M.~P.~Solon,
  Phys.\ Rev.\ D {\bf 91}, 043505 (2015) [arXiv:1409.8290
  [hep-ph]]. See, in particular, Table~10.

\bibitem{Junnarkar:2013ac} 
  P.~Junnarkar and A.~Walker-Loud,
  Phys.\ Rev.\ D {\bf 87}, 114510 (2013)
  [arXiv:1301.1114 [hep-lat]].




\bibitem{Anchordoqui:2013pta} 
  L.~A.~Anchordoqui and B.~J.~Vlcek,
  Phys.\ Rev.\ D {\bf 88}, 043513 (2013)
  [arXiv:1305.4625 [hep-ph]].


\bibitem{Agashe:2014kda} 
  K.~A.~Olive {\it et al.}  [Particle Data Group Collaboration],
  Chin.\ Phys.\ C {\bf 38}, 090001 (2014).

\bibitem{Garcia-Cely:2013nin} 
  C.~Garcia-Cely, A.~Ibarra and E.~Molinaro,
  JCAP {\bf 1311}, 061 (2013)
  [arXiv:1310.6256 [hep-ph]].

\bibitem{Gondolo:1990dk} 
  P.~Gondolo and G.~Gelmini,
  Nucl.\ Phys.\ B {\bf 360}, 145 (1991).



\bibitem{Griest:1989wd} 
  K.~Griest and M.~Kamionkowski,
  Phys.\ Rev.\ Lett.\  {\bf 64}, 615 (1990).


\bibitem{Blum:2014dca} 
  K.~Blum, Y.~Cui and M.~Kamionkowski,
  arXiv:1412.3463 [hep-ph].


\bibitem{Akerib:2013tjd} 
  D.~S.~Akerib {\it et al.}  [LUX Collaboration],
  Phys.\ Rev.\ Lett.\  {\bf 112}, 091303 (2014)
  [arXiv:1310.8214 [astro-ph.CO]].


\bibitem{Agnese:2014aze} 
  R.~Agnese {\it et al.}  [SuperCDMS Collaboration],
  Phys.\ Rev.\ Lett.\  {\bf 112},  241302 (2014)
  [arXiv:1402.7137 [hep-ex]].


\bibitem{Choi:2015ara} 
  K.~Choi {\it et al.}  [Super-Kamiokande Collaboration],
  Phys.\ Rev.\ Lett.\  {\bf 114},  141301 (2015)
  [arXiv:1503.04858 [hep-ex]].



\bibitem{Schmidt-Hoberg:2013hba} 
  K.~Schmidt-Hoberg, F.~Staub and M.~W.~Winkler,
  Phys.\ Lett.\ B {\bf 727}, 506 (2013)
  [arXiv:1310.6752 [hep-ph]].

\bibitem{Clarke:2013aya} 
  J.~D.~Clarke, R.~Foot and R.~R.~Volkas,
  JHEP {\bf 1402}, 123 (2014)
  [arXiv:1310.8042 [hep-ph]].


\bibitem{Anchordoqui:2013bfa} 
  L.~A.~Anchordoqui, P.~B.~Denton, H.~Goldberg, T.~C.~Paul, L.~H.~M.~Da Silva, B.~J.~Vlcek and T.~J.~Weiler,
  Phys.\ Rev.\ D {\bf 89},  083513 (2014)
  [arXiv:1312.2547 [hep-ph]].




\bibitem{delAmoSanchez:2010ac} 
  P.~del Amo Sanchez {\it et al.}  [BaBar Collaboration],
  Phys.\ Rev.\ Lett.\  {\bf 107}, 021804 (2011)
  [arXiv:1007.4646 [hep-ex]].




\bibitem{Huang:2013oua} 
  F.~P.~Huang, C.~S.~Li, D.~Y.~Shao and J.~Wang,
  arXiv:1307.7458 [hep-ph].



\bibitem{Aubert:2008as} 
  B.~Aubert {\it et al.}  [BaBar Collaboration],
  arXiv:0808.0017 [hep-ex].


\bibitem{Insler:2010jw} 
  J.~Insler {\it et al.}  [CLEO Collaboration],
  Phys.\ Rev.\ D {\bf 81}, 091101 (2010)
  [arXiv:1003.0417 [hep-ex]].



\bibitem{Buskulic:1993gi} 
  D.~Buskulic {\it et al.}  [ALEPH Collaboration],
  Phys.\ Lett.\ B {\bf 313}, 312 (1993).



\bibitem{Acciarri:1996um} 
  M.~Acciarri {\it et al.}  [L3 Collaboration],
  Phys.\ Lett.\ B {\bf 385}, 454 (1996).



\bibitem{Alexander:1996ne} 
  G.~Alexander {\it et al.}  [OPAL Collaboration],
  Phys.\ Lett.\ B {\bf 377}, 273 (1996).


\bibitem{Abbiendi:2002qp} 
  G.~Abbiendi {\it et al.}  [OPAL Collaboration],
  Eur.\ Phys.\ J.\ C {\bf 27}, 311 (2003)
  [hep-ex/0206022].

\bibitem{Abbiendi:2007ac} 
  G.~Abbiendi {\it et al.}  [OPAL Collaboration],
  Phys.\ Lett.\ B {\bf 682}, 381 (2010)
  [arXiv:0707.0373 [hep-ex]].

\bibitem{Cheung:2013oya} 
  K.~Cheung, W.~Y.~Keung and T.~C.~Yuan,
  Phys.\ Rev.\ D {\bf 89}, 015007 (2014)
  [arXiv:1308.4235 [hep-ph]].


\bibitem{Aubert:2004ws}  
  B.~Aubert {\it et al.}  [BaBar Collaboration],
  Phys.\ Rev.\ Lett.\  {\bf 94}, 101801 (2005)
  [hep-ex/0411061].

\bibitem{delAmoSanchez:2010bk} 
  P.~del Amo Sanchez {\it et al.}  [BaBar Collaboration],
  Phys.\ Rev.\ D {\bf 82}, 112002 (2010)
  [arXiv:1009.1529 [hep-ex]].

\bibitem{Lees:2013kla} 
  J.~P.~Lees {\it et al.}  [BaBar Collaboration],
  Phys.\ Rev.\ D {\bf 87}, 112005 (2013)
  [arXiv:1303.7465 [hep-ex]].


\bibitem{Browder:2000qr} 
  T.~E.~Browder {\it et al.}  [CLEO Collaboration],
  Phys.\ Rev.\ Lett.\  {\bf 86}, 2950 (2001)
  [hep-ex/0007057].



\bibitem{Lutz:2013ftz} 
  O.~Lutz {\it et al.}  [Belle Collaboration],
  Phys.\ Rev.\ D {\bf 87}, 111103 (2013)
  [arXiv:1303.3719 [hep-ex]].



\bibitem{Adler:2001xv} 
  S.~Adler {\it et al.}  [E787 Collaboration],
  Phys.\ Rev.\ Lett.\  {\bf 88}, 041803 (2002)
  [hep-ex/0111091].


\bibitem{Anisimovsky:2004hr} 
  V.~V.~Anisimovsky {\it et al.}  [E949 Collaboration],
  Phys.\ Rev.\ Lett.\  {\bf 93}, 031801 (2004)
  [hep-ex/0403036].



\bibitem{Adler:2008zza} 
  S.~Adler {\it et al.}  [E949 and E787 Collaborations],
  Phys.\ Rev.\ D {\bf 77}, 052003 (2008)
  [arXiv:0709.1000 [hep-ex]].





\bibitem{Artamonov:2009sz} 
  A.~V.~Artamonov {\it et al.}  [BNL-E949 Collaboration],
  Phys.\ Rev.\ D {\bf 79}, 092004 (2009)
  [arXiv:0903.0030 [hep-ex]].



\bibitem{Barger:2012hv} 
  V.~Barger, M.~Ishida and W.~Y.~Keung,
  Phys.\ Rev.\ Lett.\  {\bf 108}, 261801 (2012)
  [arXiv:1203.3456 [hep-ph]].


\bibitem{Espinosa:2012vu} 
  J.~R.~Espinosa, M.~Muhlleitner, C.~Grojean and M.~Trott,
  JHEP {\bf 1209}, 126 (2012)
  [arXiv:1205.6790 [hep-ph]].


\bibitem{Cheung:2013kla} 
  K.~Cheung, J.~S.~Lee and P.~-Y.~Tseng,
  JHEP {\bf 1305}, 134 (2013)
  [arXiv:1302.3794 [hep-ph]].

\bibitem{Giardino:2013bma} 
  P.~P.~Giardino, K.~Kannike, I.~Masina, M.~Raidal and A.~Strumia,
  JHEP {\bf 1405}, 046 (2014)
  [arXiv:1303.3570 [hep-ph]].

\bibitem{Ellis:2013lra} 
  J.~Ellis and T.~You,
  JHEP {\bf 1306}, 103 (2013)
  [arXiv:1303.3879 [hep-ph]].




\bibitem{Raffelt:1990yz} 
  G.~G.~Raffelt,
  Phys.\ Rept.\  {\bf 198}, 1 (1990).

\bibitem{Keung:2013mfa} 
  W.~Y.~Keung, K.~W.~Ng, H.~Tu and T.~C.~Yuan,
  Phys.\ Rev.\ D {\bf 90},  075014 (2014)
  [arXiv:1312.3488 [hep-ph]].

\bibitem{Khachatryan:2014rra} 
  V.~Khachatryan {\it et al.} [CMS Collaboration],
  Eur.\ Phys.\ J.\ C {\bf 75}, no. 5, 235 (2015)
  [arXiv:1408.3583 [hep-ex]].

\bibitem{LopezHonorez:2012kv} 
  L.~Lopez-Honorez, T.~Schwetz and J.~Zupan,
  Phys.\ Lett.\ B {\bf 716}, 179 (2012)
  [arXiv:1203.2064 [hep-ph]].



\bibitem{Carpenter:2013xra} 
  L.~Carpenter, A.~DiFranzo, M.~Mulhearn, C.~Shimmin, S.~Tulin and D.~Whiteson,
  Phys.\ Rev.\ D {\bf 89},  075017 (2014)
  [arXiv:1312.2592 [hep-ph]].


\bibitem{Abdallah:2015wta} 
  J.~Abdallah {\it et al.},
  arXiv:1506.03116 [hep-ph].





\bibitem{Steigman:1977kc} 
  G.~Steigman, D.N.~Schramm and J.E.~Gunn,
  Phys.\ Lett.\ B {\bf 66}, 202 (1977).


\bibitem{Ade:2015xua} 
  P.~A.~R.~Ade {\it et al.}  [Planck Collaboration],
  arXiv:1502.01589 [astro-ph.CO].


\bibitem{Mangano:2005cc} 
  G.~Mangano, G.~Miele, S.~Pastor, T.~Pinto, O.~Pisanti and P.~D.~Serpico,
  Nucl.\ Phys.\ B {\bf 729}, 221 (2005)
  [hep-ph/0506164].

\bibitem{Aver:2013wba} 
  E.~Aver, K.~A.~Olive, R.~L.~Porter and E.~D.~Skillman,
  JCAP {\bf 1311}, 017 (2013)
  [arXiv:1309.0047 [astro-ph.CO]].

\bibitem{Cooke:2013cba} 
  R.~Cooke, M.~Pettini, R.~A.~Jorgenson, M.~T.~Murphy and C.~C.~Steidel,
  arXiv:1308.3240 [astro-ph.CO].


\bibitem{Nollett:2014lwa} 
  K.~M.~Nollett and G.~Steigman,
  Phys.\ Rev.\ D {\bf 91},  083505 (2015)
  [arXiv:1411.6005 [astro-ph.CO]].

\bibitem{Arason:1991ic} 
  H.~Arason, D.~J.~Castano, B.~Keszthelyi, S.~Mikaelian, E.~J.~Piard, P.~Ramond and B.~D.~Wright,
  Phys.\ Rev.\ D {\bf 46}, 3945 (1992).


\bibitem{Iso:2009ss} 
  S.~Iso, N.~Okada and Y.~Orikasa,
  Phys.\ Lett.\ B {\bf 676}, 81 (2009)
  [arXiv:0902.4050 [hep-ph]].




\bibitem{Casas:1994us} 
  J.~A.~Casas, J.~R.~Espinosa, M.~Quiros and A.~Riotto,
  Nucl.\ Phys.\ B {\bf 436}, 3 (1995)
  [Erratum-ibid.\ B {\bf 439}, 466 (1995)]
  [hep-ph/9407389].


\bibitem{Weinberg:1976pe} 
  S.~Weinberg,
  Phys.\ Rev.\ Lett.\  {\bf 36}, 294 (1976).



\bibitem{Linde:1975gx} 
  A.~D.~Linde,
  Phys.\ Lett.\ B {\bf 62}, 435 (1976).


\bibitem{Sher:1993mf} 
  M.~Sher,
  Phys.\ Lett.\ B {\bf 317}, 159 (1993)
  [Phys.\ Lett.\ B {\bf 331}, 448 (1994)]
  [hep-ph/9307342].



\bibitem{Aad:2014yja} 
  G.~Aad {\it et al.}  [ATLAS Collaboration],
  Phys.\ Rev.\ Lett.\  {\bf 114},  081802 (2015)
  [arXiv:1406.5053 [hep-ex]].

\bibitem{Khachatryan:2015yea} 
  V.~Khachatryan {\it et al.}  [CMS Collaboration],
  arXiv:1503.04114 [hep-ex].



\bibitem{Feng:2014uja} 
  J.~L.~Feng {\it et al.},
  arXiv:1401.6085 [hep-ex].


\bibitem{Baudis:2014naa} 
  L.~Baudis,
  Phys.\ Dark Univ.\  {\bf 4}, 50 (2014)
  [arXiv:1408.4371 [astro-ph.IM]].

\end{thebibliography}
\end{document}